\documentclass[12pt]{iopart}

\usepackage{iopams}
\usepackage{verbatim}
\usepackage{graphicx}
\usepackage{color}
 \newcommand{\ket}[1]{\left|#1\right\rangle}

 \newcommand{\text}{\textnormal} 
 
\usepackage{color}

\usepackage{umoline}

\usepackage[numbers,sort&compress]{natbib}

\begin{document}


\title{U(1) Wilson lattice gauge theories in digital quantum simulators}

\author{Christine Muschik$^{1,2}$, Markus Heyl$^{2,3}$, Esteban Martinez$^{4}$, Thomas Monz$^{4}$,  Philipp Schindler$^{4}$, Berit Vogell$^{1,2}$, Marcello Dalmonte$^{1,5}$, Philipp Hauke$^{1,2}$, Rainer Blatt$^{4,6}$, Peter Zoller$^{1,6}$ }

\address{$^1$Institute for Theoretical Physics, University of Innsbruck, A-6020, Innsbruck, Austria}
\address{$^2$Institute for Quantum Optics and Quantum Information of the Austrian Academy of Sciences, A-6020 Innsbruck, Austria}
\address{$^2$Department of Physics, Walter Schottky Institute, and Institute for Advanced Study,
Technical University of Munich, 85748 Garching, Germany}
\address{$^3$Max-Planck-Institut fuer Physik komplexer Systeme, 01187 Dresden, Germany}
\address{$^4$Institute for Experimental Physics, University of Innsbruck, 6020 Innsbruck, Austria}
\address{$^5$Abdus Salam International Center for Theoretical Physics, Strada Costiera 11, Trieste, Italy}

\begin{abstract}
Lattice gauge theories describe fundamental phenomena in nature, but calculating their real-time dynamics on classical computers is notoriously difficult. In a recent publication [Nature 534, 516 (2016)], we proposed and experimentally demonstrated a digital quantum simulation of the paradigmatic Schwinger model, a U(1)-Wilson lattice gauge theory describing the interplay between fermionic matter and gauge bosons. Here, we provide a detailed theoretical analysis of the performance and the potential of this protocol. Our strategy is based on analytically integrating out the gauge bosons, which preserves exact gauge invariance but results in complicated long-range interactions between the matter fields. Trapped-ion platforms are naturally suited to implementing these interactions, allowing for an efficient quantum simulation of the model, with a number of gate operations that scales only polynomially with system size. Employing numerical simulations, we illustrate that relevant phenomena can be observed in larger experimental systems, using as an example the production of particle--antiparticle pairs after a quantum quench. We investigate theoretically the robustness of the scheme towards generic error sources, and show that near-future experiments can reach regimes where finite-size effects are insignificant. We also discuss the challenges in quantum simulating the continuum limit of the theory. Using our scheme, fundamental phenomena of lattice gauge theories can be probed using a broad set of experimentally accessible observables, including the entanglement entropy and the vacuum persistence amplitude. 
\end{abstract}

\maketitle

\section{Introduction}\label{Sec_Introduction}
In~\cite{ExpPaper}, we presented an efficient scheme that allows for the quantum simulation of real-time dynamics of lattice gauge theories and reported on its experimental realisation in a system of trapped ions. 
The purpose of the present paper is twofold: first, we give a detailed account of the theoretical basis behind the experimental demonstration in~\cite{ExpPaper}. Second, we address questions relevant to future experimental work, including a careful analysis of the effect of imperfections and a discussion of the scalability of the approach to mesoscopic system sizes.
We concentrate here on trapped-ion implementations of the type described in~\cite{Blatt2012,Schindler2013}. This platform allows one to realize a spin chain where (i)  local operations can be performed with high fidelity and (ii) an all-to-all two-body interaction can be induced between the spin degrees of freedom using so-called M\o lmer-S\o rensen gates~\cite{MSgates,Milburn99,Solano99,Leibfried2003}. Our scheme uses these resources in an highly efficient way and is specifically tailored to realize Wilson's formulation of gauge theories on a discrete lattice, which provides an ideal starting point to investigate the dynamics of gauge theories within a non-perturbative framework~\cite{Wilson74,Montvay1994,Gattringer2010}. 

At equilibrium, and in certain parameter regimes (e.g. at zero chemical potential) quantum Monte Carlo simulations of lattice gauge theories can be carried out very efficiently~\cite{Montvay1994,Gattringer2010}. However, non-equilibrium properties, as relevant for a variety of high-energy physics phenomena including particle--antiparticle production at high-intensity laser facilities~\cite{Eli,Narozhny2014}, are not accessible, due to the fundamental sign (or complex action) problem affecting simulations in real time~\cite{calzetta_book}. In the last few years, several proposals for quantum simulations of real-time dynamics of lattice gauge theories have been put forward~\cite{Wiese:2013kk,Zohar2015,Dalmonte:2016jk}, based on a variety of platforms ranging from cold atoms in optical lattices~\cite{Zohar2012,Banerjee2012,Tagliacozzo:2012kq,Glaetzle:2014dq,Bazavov:2015ly,Notarnicola:2015db,Kasper2015}, to superconducting circuits~\cite{Marcos:2013rz,Egusquiza2015} and trapped ions~\cite{Hauke2013b,Yang2016}. The main difficulties in implementing gauge theories in quantum simulators stem from the fact that complex many-body interactions have to be realized, while at the same time local (gauge) symmetries have to be imposed on the system dynamics~\cite{Wiese:2013kk,Zohar2015,Dalmonte:2016jk}. 
To address these challenges, we proposed~\cite{ExpPaper} to use encoding techniques, which exploit an analytical elimination of gauge fields. This approach allowed us to realize a digital quantum simulation scheme~\cite{Lloyd96} in a system of trapped ions, that simulates the Schwinger model~\cite{Schwinger1,Schwinger1951}, which describes quantum electrodynamics in (1+1) dimensions (1 spatial dimension + time).

The Schwinger model represents a simple, yet paradigmatic example of a U(1) gauge theory. It describes the coupling of fermions to a dynamical electromagnetic field in (1+1) dimensions, and exhibits a series of phenomena, such as chiral symmetry breaking and confinement~\cite{calzetta_book,Coleman1976239}, that play a key role in the current understanding of more complex theories such as quantum chromodynamics~\cite{Gattringer2010}. The real-time dynamics of the Schwinger model includes the spontaneous creation of particle--antiparticle pairs~\cite{Hebenstreit2013}. Despite the simplicity of the model, such phenomena are notoriously hard to address using numerical simulations, mostly due to the absence of unbiased methods~\footnote{In parameter regimes characterized by high occupancies of the photon degrees of freedom, semiclassical simulations have recently been shown to provide accurate results~\cite{Kasper:2014xe}, while tensor-network methods~\cite{Wiese:2013kk,Banuls,Rico:2014ek,Buyens:2014cs} have been successfully applied for short-time dynamics~\cite{Pichler:2016it,Buyens:2015lq}.}. The Schwinger model provides a good starting point for studies of lattice gauge theories, since interesting physical insights can be gained using moderate experimental resources due to the reduced dimensionality. In quantum simulators, the real-time dynamics can be probed using a rich set of observables including entanglement entropies and vacuum persistence amplitudes. While the study of quantum information concepts such as entanglement in the context of high-energy physics is a rather recent development~\cite{Calabrese:2006pb,Nishioka:2009sf,Casini2014}, vacuum persistence amplitudes play an important role for the theoretical understanding of spontaneous pair creation already in Schwinger's original work \cite{Schwinger1,Schwinger2}. \\

\begin{figure}[t]
\centering
\includegraphics[width=\columnwidth, angle=0]{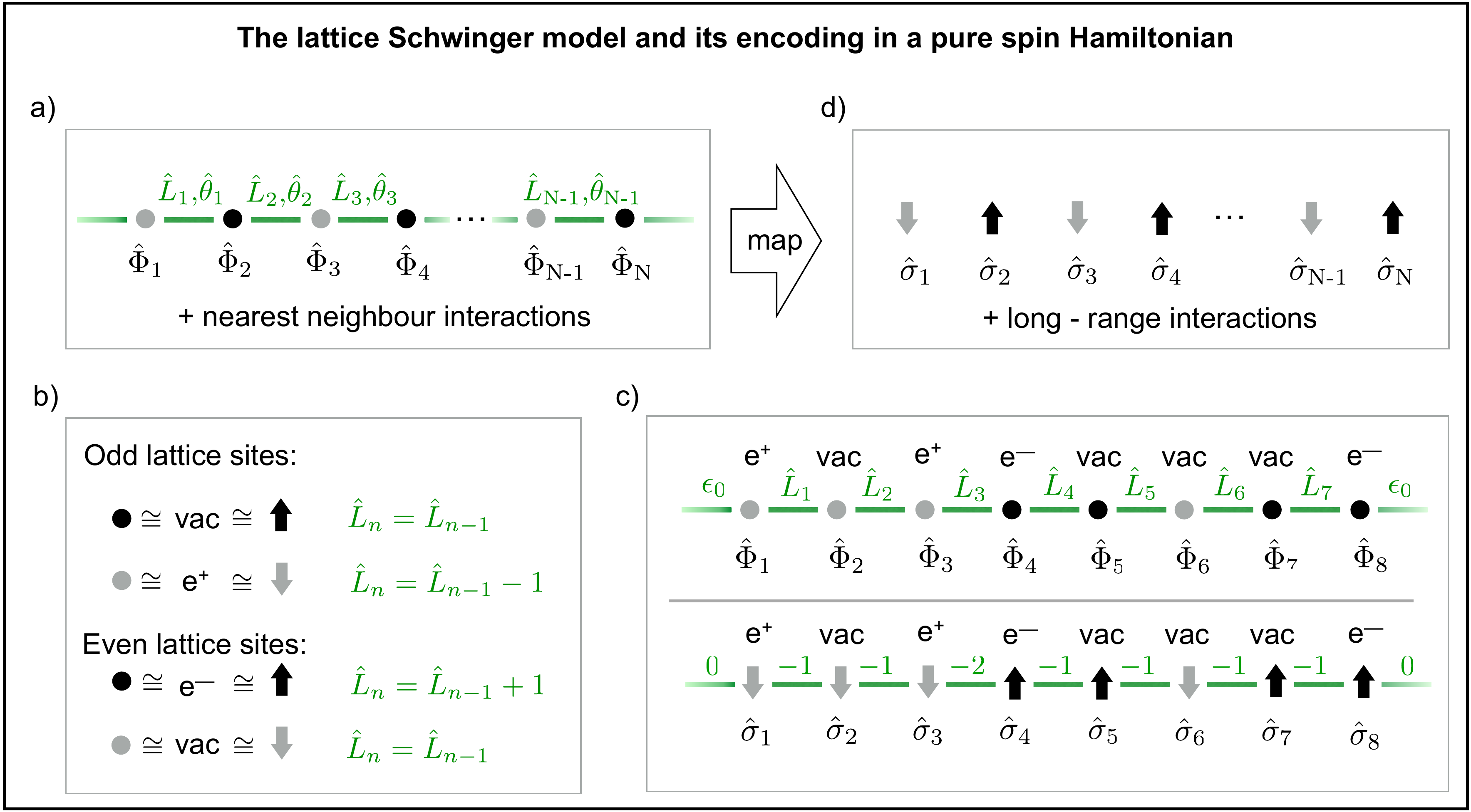}\label{Fig_1}
\caption{\label{Fig_StaggeredFermions} Encoding of the lattice Schwinger model. 
(a) Matter fields are represented by one-component fermion fields $\hat{\Phi}_n$ at lattice sites $n$ that couple via gauge variables $\hat{L}_n$ (electric fields) and $\hat{\theta}_n$ (vector potentials) defined on the links between sites. The interaction is governed by the Schwinger lattice Hamiltonian given by Eq.~(\ref{Eq_LatticeHamiltonian}) in the main text.
(b) Translation table: occupied even (unoccupied odd) lattice sites translate to the presence of an electron (positron). Unoccupied even and occupied odd lattice sites represent the vacuum. The translation is analogous after mapping the fermion fields to spins. For each configuration, the Gauss law enforces a relation between the adjacent gauge fields as depicted in green. 
(c) For illustration, a matter configuration with eight lattice sites is shown in the lattice Schwinger model (upper panel) and after mapping the fermion fields $\hat{\Phi}_n$ to Pauli spin operators $\hat{\sigma}_n$
(lower panel). Due to the Gauss law, the gauge fields are completely determined for a given matter configuration and choice of background field.
(d) The Gauss law allows for the elimination of the gauge degrees of freedom. If the fermion fields  are mapped to Pauli spin operators, the system Hamiltonian becomes a pure spin model with long-range interactions, which correspond to the Coulomb interaction between the simulated charged particles. 
}
\end{figure}

In this article, we explain in detail how an efficient implementation of the Schwinger model can be carried out by combining the toolkit of digital quantum simulation  \cite{Leibfried2003,Blatt2012,Schindler2013,Monroe2013,Barends15} with techniques used in numerical computations on classical computers (so-called encoding techniques~\cite{Encoding}). In this model of (1+1)d quantum electrodynamics, electrons and positrons appear in the form of fermion fields that are defined on a lattice, and which interact via electric-field gauge bosons that are defined on the links between lattice sites (see Fig.~\ref{Fig_StaggeredFermions}). 
For implementing the model, the fermionic operators can be conveniently mapped to Pauli spin operators, which are the natural degrees of freedom in spin-based quantum simulators \cite{Leibfried2003,Blatt2012,Schindler2013,Monroe2013}. The gauge field operators, in contrast, are associated with an infinite-dimensional Hilbert space, which poses a challenge for their implementation on a quantum simulator with bounded Hilbert space \cite{Wiese:2013kk}. 
To circumvent this difficulty, several proposals have considered quantum link models~\cite{QLink1,QLink2,QLink3,QLink4}, where the gauge Hilbert space is truncated and gauge fields are represented by spin-operators of finite dimensionality. Here, we follow an alternative approach and realize Wilson's formulation of lattice gauge theories~\cite{Wilson74}, which takes the full infinite-dimensional Hilbert space of the gauge degrees of freedom into account.

As explained in detail below, our simulation scheme relies on analytically integrating out the gauge fields~\cite{Encoding}, which is reminiscent of the analytical elimination of the fermionic fields in Monte Carlo methods~\cite{Gattringer2010}. 
In this way, we obtain an effective description in terms of a pure spin model where the gauge fields no longer appear explicitly but rather enter in the form of long-range interactions with an exotic distance dependence. 
Previously, this approach has been employed to aid analytical or numerical calculations~\cite{Encoding,MC1,MC2,Saito:2015ai,Banuls:2016qe}. In contrast, we use this idea for quantum simulation, i.e.\ the realisation of the Schwinger model in its encoded form in an actual physical system. A key difficulty in realising such a simulation is the anisotropic form of the long-range interactions, as illustrated in Fig.~\ref{Fig_Hzz}. Below, we will show how these interactions can be implemented efficiently in a digital quantum simulator that features single-qubit operations and entangling gates between arbitrary pairs of spins. These resources are naturally available in trapped-ion setups, where these gate operations can be performed with high accuracy  \cite{Leibfried2003,Blatt2012,Schindler2013,Monroe2013}. This platform therefore provides an ideal match for the realisation of the proposed scheme.

As mentioned above, one of the key challenges for quantum simulations of gauge theories~\cite{Preskill2012,Wiese:2013kk,Zohar2015} is the requirement that the dynamics must obey gauge invariance, i.e.\  they must take place in the subspace corresponding to states that are physically allowed in the simulated model. This requirement translates into local constraints that govern the interaction between matter and gauge fields. In the case of quantum electrodynamics, the constraints are imposed by the Gauss law, which in the continuum limit is given by $\nabla E=\rho$, where $E$ is the electric field and $\rho$ is the charge density. In quantum simulation proposals where both matter and gauge fields are explicitly present, gauge invariance is typically imposed by suppressing processes that take the dynamics out of the allowed subspace~\cite{Wiese:2013kk}, for example by enforcing energy penalties. Hence, the resulting dynamics is only gauge invariant up to some energy scale. Our approach has the advantage that, by construction, the dynamics takes place in the subspace where gauge invariance is automatically fulfilled. Instead of introducing $2N-1$ quantum systems to simulate $N$ particles together with the accompanying gauge fields and restricting the dynamics to a much smaller Hilbert space, in this approach we simulate the dynamics of $N$ fermions using only $N$ spins. The combination of the mapping to a pure spin model and its realisation by means of a digital simulation scheme allows therefore for a very efficient use of resources, which renders the quantum simulation of the Schwinger model possible with present-day experimental means.\\

The remainder of this paper is organized as follows. Section~\ref{Sec_Scheme} presents the model under consideration and discusses the details of our quantum-simulation scheme. We explain the encoding strategy, and describe how the resulting highly non-local Hamiltonian can be realized efficiently in a digital quantum simulator.  
In section~\ref{Sec_SchwingerMechanism}, we illustrate the capabilities of our approach by showing how it can be used to simulate the spontaneous generation of particle--antiparticle pairs out of the vacuum of the bare fermionic particles. 
Here, we explain how the resulting decay of the vacuum can be monitored by studying the vacuum persistence amplitude and how the creation of entanglement during the pair creation process can be studied. 
In section~\ref{Sec_Implementation}, we discuss concrete implementations with ions confined in a linear Paul trap, as has recently been reported in~\cite{ExpPaper}. In particular, we analyse the effects of imperfections and discuss the scalability of the approach. In section~\ref{Sec_ContinuumLimit}, we discuss the challenges in taking the continuum limit of the theory. Finally, in section~\ref{Sec_Conclusions}, we present our conclusions and an outlook. 
%
%
%
\section{Digital quantum simulation of the Schwinger model}\label{Sec_Scheme}
In this section, we introduce the model under consideration and explain how it can be mapped to a pure spin Hamiltonian with long-range interactions by eliminating the gauge fields exactly~\cite{Encoding} (section~\ref{SubSec_Encoding}). Afterwards, we describe how the resulting spin model can be realized efficiently by means of a digital quantum simulation scheme (section~\ref{SubSec_Protocol}).
\subsection{Mapping of the Schwinger model to a spin Hamiltonian with long-range interactions}\label{SubSec_Encoding}
We consider the Schwinger model, which describes the interaction between electrons and positrons via electromagnetic fields in one spatial dimension. In the following, we give an overview to this model in the continuum and on a lattice following the description in~\cite{Encoding}. To this end, we introduce the vector potential at position $x$ with temporal and spatial component $\hat{A}_0(x)$ and $\hat{A}_1(x)$. In one spatial dimension, the electric field has only one component 
 $\hat{E}(x)=-\partial_0{\hat{A}}_1(x)$, where $\partial_0$ is the partial derivative with respect to time. $\hat{E}(x)$ represents the canonical momentum conjugate to $\hat{A}_1(x)$ with $[\hat{A}_1(x),\hat{E}(x')]=-i\delta(x-x')$. Matter fields are represented by two-component spinor fields ${\Psi}(x)=\left(\hat{\Psi}_{e^{-}}(x),\hat{\Psi}^{\dag}_{e^+}(x)\right)^{T}$. In the temporal gauge, the Schwinger Hamiltonian in the continuum is given by
\begin{eqnarray*}
\hat{H}_{\text{\tiny{cont}}}=\int dx \left[-i\bar{\Psi}(x)\gamma^1\left(\partial_1+ig\hat{A}_1(x)\right){\Psi}(x)+m\bar{\Psi}(x)\Psi(x)+\frac{1}{2}\hat{E}^2(x)\right],
\end{eqnarray*}
where $\partial_1$ is the partial derivative with respect to $x$, $m$ is the fermion mass and $\bar{\Psi}\equiv\Psi^{\dag}\gamma^0$. In one spatial dimension, the Dirac matrices $\gamma^{0}$ and $\gamma^1$ are given by the Pauli operators $\gamma^0=\hat{\sigma}^z$ and $\gamma^1=i\hat{\sigma}^y$. Using natural units $\hbar=c=1$, the coupling constant $g=-e$ is given by the charge $e$ of the elementary particles. This model can be formulated on a lattice where points in space are separated by a distance $a$, while time is continuous. In the following, we are using the so-called Kogut-Susskind Hamiltonian formulation~\cite{Schwinger1,Schwinger2,KogutSusskindFormulation} of the lattice Schwinger model. In this description, the continuous fields $\hat{A}_1(x)$ and $\hat{E}(x)$ are replaced by conjugate variables $\hat{\theta}_n=-a g \hat{A}_1(x_n)$ and $\hat{L}_n=\frac{1}{g}\hat{E}(x_n)$. The operators $\hat{L}_n$, $\hat{\theta}_n$ are defined on the links connecting lattice sites $n$ and $n+1$ as shown in Fig.~\ref{Fig_1}(a) and commute canonically $[\hat{\theta}_n,\hat{L}_m]=i\delta_{n,m}$. Particles are represented by Kogut-Susskind fermions, with one-component fermion field operators defined on each site $n$: $\hat{\Phi}_n=\sqrt{a}\hat{\Psi}_{e^{-}}(x_n)$ for even $n$ and $\hat{\Phi}_n=\sqrt{a}\hat{\Psi}^{\dag}_{e^{+}}(x_n)$ for odd $n$. The unit cell of this staggered lattice consists therefore of two sites and the presence of an electron (positron) is indicated by an occupied even (unoccupied odd) site, as sketched in Fig.~\ref{Fig_1}(b).
Accordingly, the interaction of matter- and gauge fields is described by the lattice Schwinger Hamiltonian
\begin{eqnarray}\label{Eq_LatticeHamiltonian}
\hat{H}_{{\rm{lat}}}&=&-i w\sum_{n=1}^{N-1}\left[\hat{\Phi}^{\dag}_ne^{i\hat{\theta}_n}\hat{\Phi}_{n+1}-H.C.\right]
+m \sum_{n=1}^{N}(-1)^n \hat{\Phi}^{\dag}_n\hat{\Phi}_n +J\sum_{n=1}^{N-1} \hat{L}_n^2,\ \  \  \ 
\end{eqnarray}
where $N$ is the number of lattice sites and $m$ is the fermion mass; $w=\frac{1}{2a}$ and $J=\frac{g^2 a}{2}$, where $a$ is the lattice constant and $g$ the fermion-light coupling constant. Using natural units $\hbar = c = 1$, the parameters $w$, $J$, $m$, and $g$ have the dimension of inverse length, while $a$ and $t$ (time) have the dimension of length.
The first term in Eq.~(\ref{Eq_LatticeHamiltonian}) describes nearest-neighbour hopping and corresponds to the creation and annihilation of electron-positron pairs~\footnote{This is illustrated in the upper half of Fig.~1(c). The fermion fields shown at lattice sites 5 and 6 represent empty space (vac,vac). A hopping process that swaps the fermion fields of these two adjacent sites leads to configuration representing an electron-positron pair, as shown at lattices sites 3 and 4.}. The second and the third term represent the rest mass and the electric field energy stored in the system, respectively. 
The resulting dynamics is constrained by the Gauss law~\footnote{In the continuum ($a \rightarrow 0$), the Gauss law is given by $\partial_1E(x)=g\bar{\Psi}\gamma^0\Psi$. In three spatial dimensions, this takes the form $\nabla E=\rho$, where $\rho$ is the charge density.}. In the considered lattice formulation, it takes the form of a set of local constraints as illustrated in Fig.~\ref{Fig_1}(b),(c). More precisely, physical states are eigenstates of the generators of the Gauss law $\hat{G}_n=\hat{L}_n-\hat{L}_{n-1} -\hat{\Phi}^{\dag}{_n}\hat{\Phi}{_n}+\frac{1}{2}\left[1-(-1)^n\right]$. We will be interested in the zero-charge subspace, where $\hat{G}_n |\psi_{\text\tiny{{physical}}}\rangle=0$. Note that the electric field operators for each link take integer eigenvalues $L_n\!=\!0,\pm1,\pm 2,...$ .\\

\begin{figure}[t]
\centering
\includegraphics[width=0.8\columnwidth, angle=0]{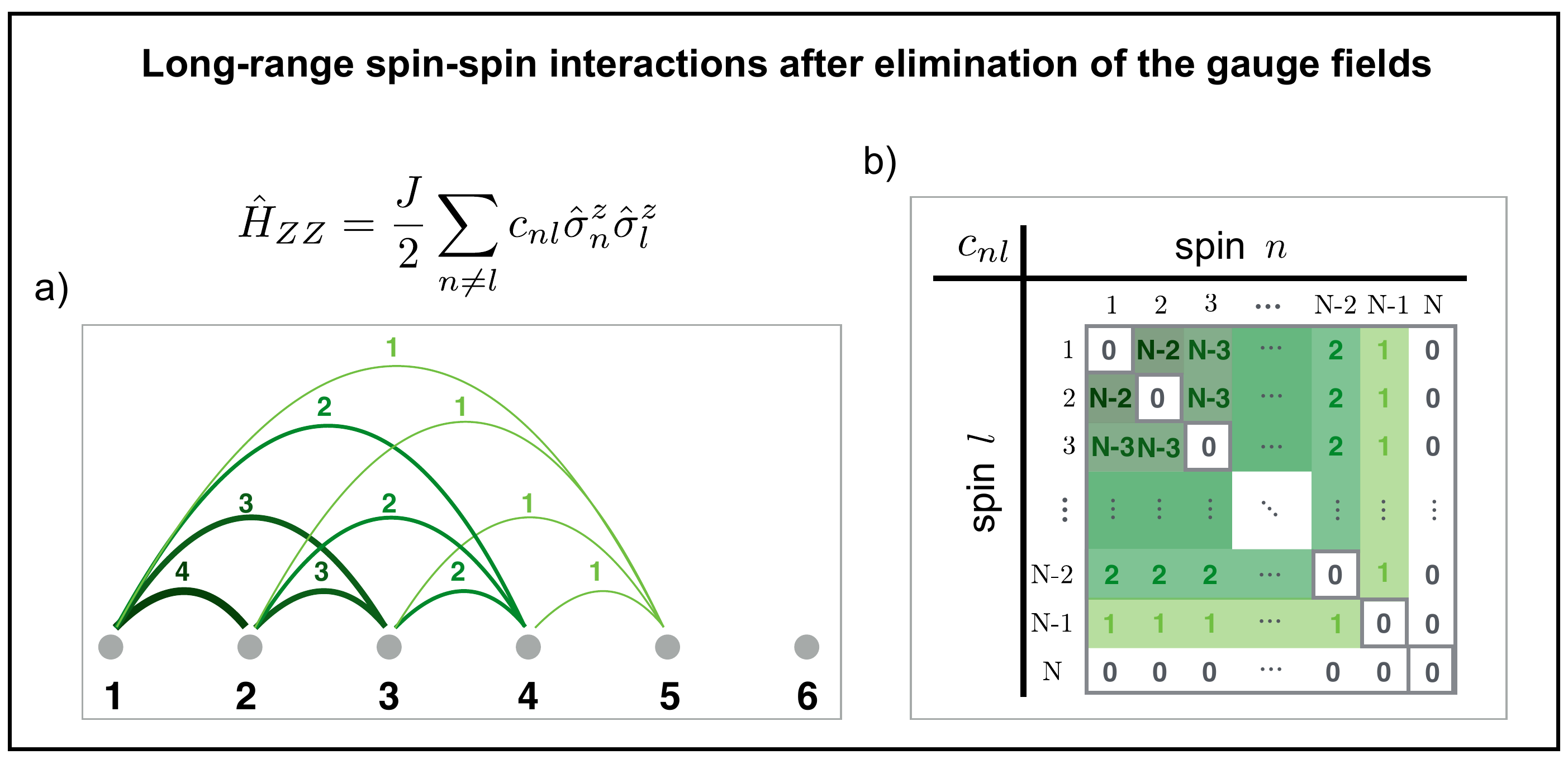}
\caption{\label{Fig_Hzz}
Eliminating the gauge fields in the Schwinger model (as described in the text and in Fig.~\ref{Fig_StaggeredFermions}) results in long-range interactions with asymmetric coupling. While every spin interacts with a constant strength with all spins to its left, the coupling to spins on its right decreases linearly with distance.
Panel (a) illustrates this asymmetric coupling for $N=6$, where the green lines denote the coupling between two particles. Number, shade, and thickness of the lines indicate  the associated interaction strength.
Panel (b) shows the full coupling matrix for the case of $N$ spins.   }
\end{figure}

We aim at realising the Schwinger model in a spin system. This can be achieved by two transformations (see~\cite{ExpPaper}, Methods section). First, the fermionic field operators $\hat{\Phi}_n$ can be mapped to Pauli spin operators by a Jordan--Wigner transformation \cite{Jordan1928},
\begin{eqnarray*}
\hat{\Phi}_n=\prod_{l<n} \left[i \hat{\sigma}_l^z\right]\hat{\sigma}_n^-.
\end{eqnarray*}
Second, the operators $\hat{\theta}_n$ can be eliminated by a gauge transformation~\cite{Encoding},
\begin{eqnarray*}
\hat{\sigma}_n^{-}\rightarrow \left[\prod_{l<n}e^{-i\hat{\theta}_l}\right]\hat{\sigma}_n^-,
\end{eqnarray*}
where each spin operator $\hat{\sigma}_n^-$ is multiplied by a phase that depends on all gauge field operators $\hat{\theta}_l$ to its left ($l<n$).
In this way, the Schwinger model can be expressed in terms of spin operators $\hat{\sigma}_n$ representing the matter fields and electric field operators $\hat{L}_n$,
\begin{eqnarray*}
\hat{H'}_{{\rm{lat}}}&=&w\sum_{n=1}^{N-1}\left[\hat{\sigma}_n^{+}\hat{\sigma}_{n+1}^{-}+H.C.\right]+\!\frac{m}{2}\sum_{n=1} ^N(-1)^n\hat{\sigma}_n^z+J\sum_{n=1}^{N-1} \hat{L}_n^2,\ \  \  \ 
\end{eqnarray*}
as shown in Fig.~\ref{Fig_1}(c) (lower panel). In this formulation, the Gauss law takes the form  $\hat{L}_n-\hat{L}_{n-1}=\frac{1}{2}\left[\hat{\sigma}_n^z+(-1)^n\right]$.
For a given spin configuration that represents a certain fermion configuration and for a given choice of background field $\epsilon_0$, the electric fields are completely determined, as  illustrated in Fig.~\ref{Fig_1}(d). More specifically, the Gauss law allows one to express the electric field operators in the form $\hat{L}_n=\epsilon_0+\frac{1}{2} \sum_{l=1}^n\left(\hat{\sigma}_l^z+(-1)^l\right)$, such that the gauge fields do no longer appear explicitly in the description~\cite{Encoding}. Instead, the electric field energy term $\hat{H}_{\rm{lat}}^{E}=J\sum_{n=1}^{N-1} \hat{L}_n^2=J\sum_{n=1}^{N-1}\left[\epsilon_0+\frac{1}{2} \sum_{l=1}^n\left(\hat{\sigma}_l^z+(-1)^l\right)\right]^2$ gives rise to (i) a long-range spin-spin interaction $\hat{H}_{ZZ}$ that corresponds to the Coulomb interaction between the charged particles and (ii) local energy offsets that lead to modified effective fermion masses. For simplicity, we will assume a zero background field (the following results can be straightforwardly generalized to arbitrary background fields). The resulting Hamiltonian can be cast in the form $\hat{H}_{{\rm{S}}}=\hat{H}_{ZZ}+\hat{H}_{\pm}+\hat{H}_Z$, with
\begin{eqnarray}
\hat{H}_{ZZ}\!&=&\!\frac{J}{2}\sum_{n=1}^{N-2}\sum_{l=n+1}^{N-1}(N-l)\hat{\sigma}_n^{z} \hat{\sigma}_l^z,\label{Eq_H_E-field}\\
\hat{H}_{\pm}\!&=&\!w\sum_{n=1}^{N-1}\left[\hat{\sigma}_n^{+}\hat{\sigma}_{n+1}^{-}+H.C.\right],\label{Eq_H_FlipFlop}\\
\hat{H}_{Z}\!&=&\!\frac{m}{2}\sum_{n=1} ^N(-1)^n\hat{\sigma}_n^z-\frac{J}{2}\sum_{n=1}^{N-1} (n{\rm{mod}}2)\sum_{l=1}^n\hat{\sigma}_l^z.\label{Eq_H_LocalTerms}
\end{eqnarray}
As outlined in the introduction, the main challenge for realising the Schwinger model in its encoded form in an actual physical system is the implementation of the long-range interaction Hamiltonian $\hat{H}_{ZZ}$, which features an exotic asymmetric distance dependence (see Fig.~\ref{Fig_Hzz}). Note that we could have encoded the electric field operators in the form $\hat{L}_n=\epsilon_0-\frac{1}{2} \sum_{l=n+1}^N\left(\hat{\sigma}_l^z+(-1)^l\right)$. This alternative encoding results in a long-range spin-spin coupling of the same type, but with reverse directionality (i.e. every spin interacts with a constant strength with all spins to its right and the coupling to spins on its left decreases linearly with distance). Both descriptions are equally valid.
%
%
\subsection{Quantum simulation protocol}\label{SubSec_Protocol}
%
\begin{figure}[tb]
\centering
\includegraphics[width=\columnwidth, angle=0]{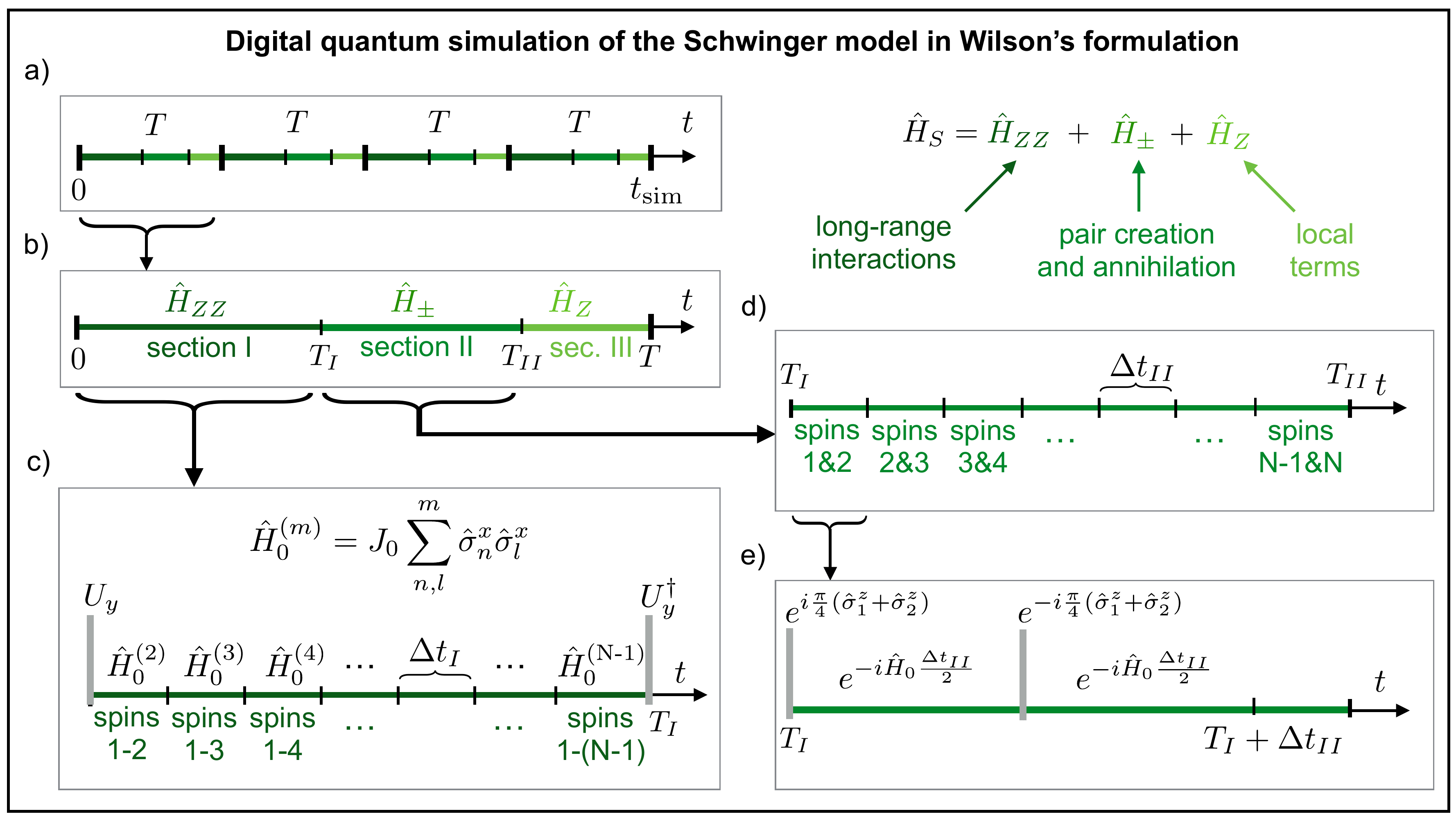}
\caption{\label{Fig_SimulationProtocol} Simulation Protocol. 
(a) The time evolution of a spin system under the Schwinger Hamiltonian $\hat{H}_{{\rm{S}}}$ is simulated by introducing discrete time steps of length $T$. 
(b) Each time window of length $T$ is divided into three sections that correspond to the three parts of the simulated Hamiltonian $\hat{H}_{ZZ}$, $\hat{H}_\pm$ and $\hat{H}_Z$, as defined in Eqs.~(\ref{Eq_H_E-field})-(\ref{Eq_H_LocalTerms}). The relative length of the three sections is not depicted to scale.  
(c) Illustration of the simulation of $\hat{H}_{ZZ}$ as given in Eq.~(\ref{Eq_H_E-field}). The first time segment (of length $T_I$) is divided into $N-2$ elementary time windows of length $\Delta t_{I}$. Within the $n^{th}$ time window, the spins interact according to the Hamiltonian $\hat{H}^{(n+1)}_0=J_0\sum_{k,l=1}^{n+1}\hat{\sigma}_k^x\hat{\sigma}_l^x$, which couples the spins $1$ to $n+1$ (while spins $n+2$ to $N$ are decoupled). To map $\hat{\sigma}^x$ to $\hat{\sigma}^z$, local rotations $U_y=e^{{i\frac{\pi}{4}}\sum_i\hat{\sigma}_i^y}$ are added at the beginning and end of this segment.
(d) Illustration of the simulation of  $\hat{H}_{\pm}$ as given in Eq.~(\ref{Eq_H_FlipFlop}). The time segment $T_{II}-T_I$ is divided into $N-1$ steps of length $\Delta t_{II}$. During each elementary time window a nearest-neighbour flip-flop interaction is realized between two selected spins, while the other spins are decoupled. 
(e) During each time window the flip-flop interaction is realized in a four step sequence as described in the main text.
}
\end{figure}
%
In the following, we describe a protocol that allows one to simulate the lattice Schwinger model in a 1D spin system with long-range interactions. The protocol consists of a digital quantum simulation scheme where we incorporated ideas put forward in~\cite{StrobShake}. Due to the complicated form of the Hamiltonian $\hat{H}_{ZZ}$ given in Eq.~(\ref{Eq_H_E-field}), a  standard digital simulation approach would require $N^2$ time steps, where $N$ is the number of spins. In our protocol, the total number of time steps scales linearly with $N$ and the realisation of  $\hat{H}_{ZZ}$ costs only $N-2$ time steps, which is optimal~\footnote{The long-range spin-spin interaction given in Eq.~(\ref{Eq_H_E-field}) is fully determined by the $N\! \times\! N$ coupling matrix shown in Fig.~\ref{Fig_Hzz}, which has a rank of $N-2$ and therefore  $N-2$ linearly independent components. This implies that the proposed simulation scheme, using only $N-2$ independent gates, is optimal.}. 
Our approach is quite general, and requires only single qubit operations that can be applied to individual spins and one type of two-body interactions, $\hat{H}_0=J_0 \sum_{n,l}\hat{\sigma}_n^a \hat{\sigma}_l^a$, where $a$ can correspond to any direction on the Bloch sphere and $J_0$ is the coupling strength. In the following, we will explain the protocol for $\hat{H_0}=J_0 \sum_{n,l}\hat{\sigma}_n^x \hat{\sigma}_l^x$. Adapting the scheme to $a\neq x$ requires only minor and straightforward modifications.\\

The simulation protocol is based on a time-coarse graining, where the effective interaction given by Eqs.~(\ref{Eq_H_E-field})-(\ref{Eq_H_LocalTerms}) is obtained in a time-averaged description, while maintaining local gauge invariance at any stage. As illustrated in Fig.~\ref{Fig_SimulationProtocol}(a), the total simulation time $t_{{\rm{sim}}}$ is divided into several time windows of duration $T$, in the spirit of the Trotter decomposition \cite{Trotter59,Lloyd96}. During each of these time windows, a full cycle of the protocol that is described below is performed. Each cycle consists of three sections as shown in Fig.~\ref{Fig_SimulationProtocol}(b). In sections I and II, $\hat{H}_{ZZ}$ and $\hat{H}_{\pm}$ are simulated employing the spin-spin coupling $\hat{H}_0$. In section III, only single particle rotations are performed realising $\hat{H}_Z$.\\

Section I is divided into $N-2$ smaller time windows of length $\Delta t_{I}$ as shown in Fig.~\ref{Fig_SimulationProtocol}(c). 
For each of the $N-2$ time windows, a subgroup of spins is decoupled from the interaction, while the remaining spins interact according to $\hat{H}_0$. 
In the $n^{{\rm{th}}}$ time window of section I, ions $1$ to $n+1$ participate in the interaction, such that the Hamiltonian $\hat{H}_0^{(n+1)}=J_0  \sum_{k,l=1}^{n+1} \hat{\sigma}_{k}^x \hat{\sigma}_{l}^x$ is implemented. 
Since the realisation of Eq.~(\ref{Eq_H_E-field}) requires $\hat{\sigma}_k^z\hat{\sigma}_l^z$ - couplings, local single-particle rotations are added in the beginning and the end of section I, 
rotating the $x$ spin component into the $z$ direction. 
The resulting time-evolution operator for section I,  
\begin{equation}
e^{{i\frac{\pi}{4}}\sum_j\hat{\sigma}_j^y}\left(\bigotimes_{m=2}^{N-1} e^{-i\hat{H}_0^{(m)}\Delta t_{I}}\right)e^{-{i\frac{\pi}{4}}\sum_j\hat{\sigma}_j^y}=e^{-i\hat{H}_{ZZ}T}\,,
\end{equation}
realizes the desired $\hat{H}_{ZZ}$ for one time step $T$, with strength $J={2}\frac{\Delta t_I}{T} J_0$.  For a single time step, this is exact. Trotter errors will be discussed in section~\ref{SubSec_TrotterErrors}, where we address imperfections of the scheme.\\

In section II, the part of the Schwinger Hamiltonian involving nearest-neighbour interactions $\hat{H}_{\pm}$ is realized. To this end, the underlying interaction $\hat{H}_0$, needs to be modified not only in range but also regarding the type of couplings. This is accomplished by dividing section II into $N-1$ elementary time slots of length $\Delta t_{II}$ (see Fig.~\ref{Fig_SimulationProtocol}(d)). Each of these time slots is used for inducing the required type of interaction between a specific pair of neighbouring atoms. This can be done by decoupling all but the selected pair of atoms from the evolution under $\hat{H}_0$. The selected pair of atoms undergoes a sequence of gate operations that transforms the $\hat{\sigma}^x_n\hat{\sigma}_l^x$-coupling into a $\left(\hat{\sigma}_n^+\hat{\sigma}_l^{-}+\hat{\sigma}_n^-\hat{\sigma}_l^{+}\right)$-interaction and consists of four steps (cf. Fig.~\ref{Fig_SimulationProtocol}(e)): (i) a single qubit operation on the two selected spins $n$ and $n+1$, $U=e^{i \frac{\pi}{4}(\hat{\sigma}^z_n+\hat{\sigma}_{n+1}^z)}$ (ii) an evolution of the qubits $n$ and $n+1$ under the Hamiltonian $\hat{H}_0$ during a time $\Delta t _{II}/2$, $e^{-i \hat{H}_0 \Delta t_{II}/2}$ (iii) another single qubit operation $U^{\dag}$ and finally (iv) another two qubit gate operation $e^{-i \hat{H}_0 \Delta t_{II}/2}$. These four steps result in a time evolution with
\begin{eqnarray*}
e^{-i\hat{H}_0 \frac{\Delta t_{II}}{2}}\cdot U^{\dag}\cdot e^{-i \hat{H}_0 \frac{\Delta t_{II}}{2}}\cdot U=e^{-i\left( \hat{H}_0+  U^{\dag}\hat{H}_0U \right) \frac{\Delta t_{II}}{2}}.
\end{eqnarray*}
Note that this equation is exact. The time evolution operator associated with the described sequence of gate operations is given by $e^{-i \hat{H}_{II} ^{(n,n+1)}\Delta t_{II}}$ with 
\begin{eqnarray*}
\hat{H}_{II}^{(n,n+1)}&=&\frac{1}{2}\left(\hat{H}_0 + U^{\dag} \hat{H}_0 U\right)=J_0\left(\hat{\sigma}_n^{+}\hat{\sigma}_{n+1}^- + H.C.\right)\,.
\end{eqnarray*}
Repeating these four steps for sites $n=1\dots N-1$ yields, as long as $J_0\Delta t_{II}(N-1)\ll 1$, the desired evolution operator $\hat{H}_{II}=\sum_{n=1}^{N-1}\hat{H}_{II}^{(n,n+1)}=\hat{H}_{\pm}$, with $w=\frac{\Delta t_{II}}{T}J_0$.
The relative strength of the nearest-neighbour Hamiltonian $\hat{H}_{\pm}$  and the $\hat{\sigma}_n^z\hat{\sigma}_l^z$-type
couplings $\hat{H}_{ZZ}$ can be adjusted by tuning the ratio of the elementary time windows $\Delta t_{II}/\Delta t_{I}$.\\
\\In section III, the single qubit terms of the type $\hat{H}_z$ given in Eq.~(\ref{Eq_H_LocalTerms})  are implemented. This Hamiltonian is realized in a single time window of length $\Delta t_{III}$. All spins are acted upon simultaneously, but each one experiences a different coupling strength. Since we assume that single qubit operations can be performed on much faster time scales than gates operating on multiple spins, we use $\Delta t_{III}\ll \Delta t_{II}, \ \Delta t_{I}$. Together, the operations in the time sections I,II, and III approximately realize the time-evolution operator of the Schwinger model for one time step, $e^{-i \hat{H}_{S}T}$. Trotter errors due to the finite-time coarse-graining will be discussed below in section~\ref{Sec_Implementation}, where we address imperfections of the scheme.
\section{Dynamics of particle production}\label{Sec_SchwingerMechanism}
%
%
\begin{figure}[t]
\centering
\includegraphics[width=\textwidth, angle=0]{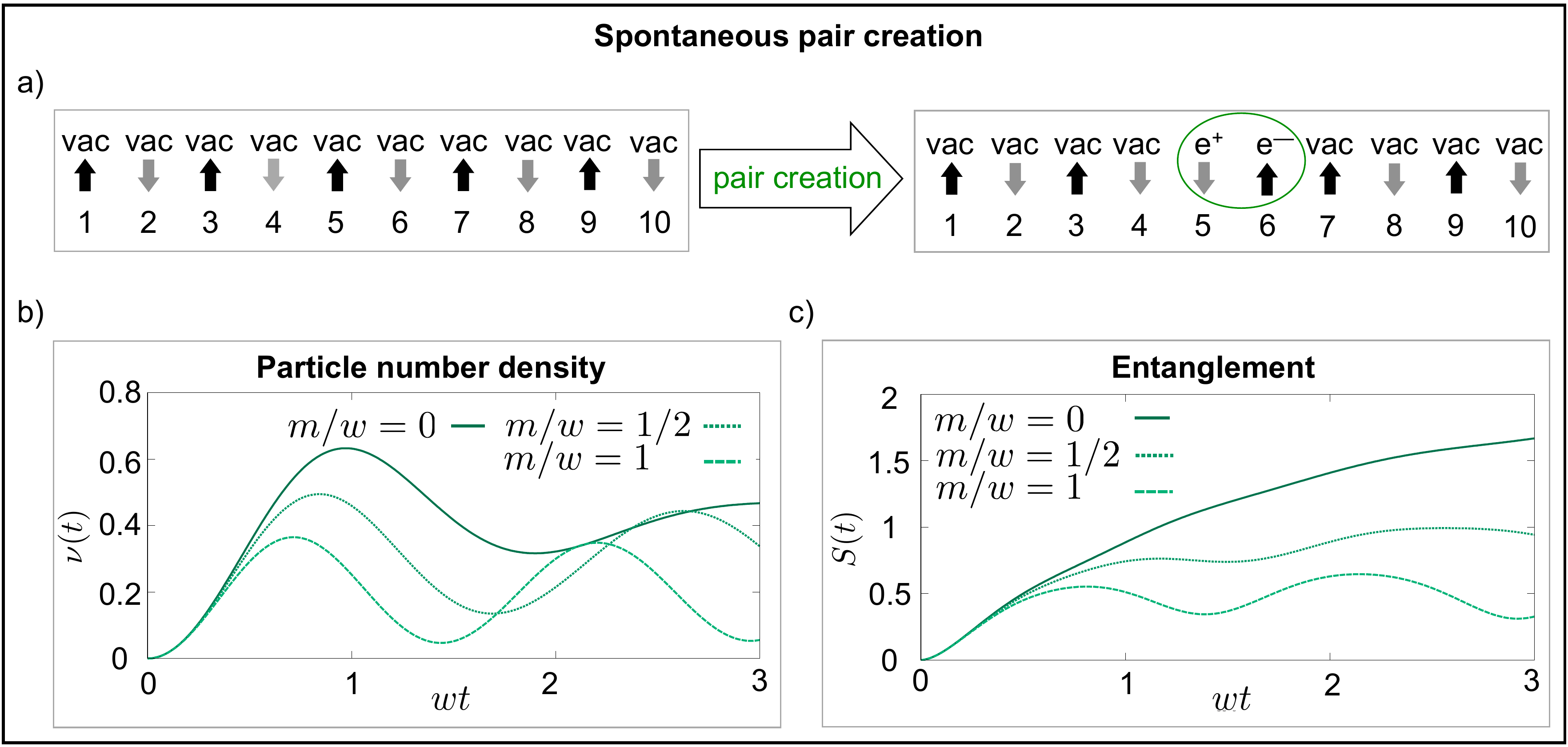}
\caption{\label{Fig_ResultsIdealCase_1} Simulation of particle production out of the bare vacuum. (a) Pair creation in the encoded Schwinger model. The left spin configuration corresponds to the bare vacuum state. The right configuration displays a state with one particle-antiparticle pair. (b,c) Instability of the bare vacuum: (b) Particle number density $\nu (t)$ and (c) entanglement entropy $S (t)$ for $J/w=1$ and different values of $m/w$, where $J$ and $w$ quantify the electric field energy and the rate at which particle-antiparticle pairs are produced, and $m$ is the fermion mass, see Eq.~(\ref{Eq_LatticeHamiltonian}). 
(b) After a fast transient pair creation regime, the increased particle density favours particle-antiparticle recombination inducing a decrease of $\nu(t)$. This nonequilibrium interplay of regimes with either dominating production or recombination continues over time and leads to an oscillatory behaviour of  $\nu(t)$ with a slowly decaying envelope. 
(c) The entanglement entropy $S(t)$ quantifies the entanglement between the left and the right half of the system, generated by the creation of particle- antiparticle pairs that are distributed across the two halves. An increasing particle mass $m$ suppresses the generation of entanglement.
}
\end{figure}
%
The proposed quantum simulation scheme allows for the experimental study of a wide range of fundamental properties in U(1)-Wilson gauge theories that are of current interest. 
For example, strong efforts are under way at high-intensity laser facilities such as ELI and XCELS to observe a cascade of particle-antiparticle pairs generated out of the vacuum subject to extreme electric fields~\cite{Eli,Narozhny2014}, and several theoretical proposals for the quantum simulation of particle production have been put forward in recent years~\cite{Wiese:2013kk,Zohar2015,Alvarez2015,Casanova2011,Kasper2016}.
The preparation of the true vacuum (the eigenstate of the Schwinger Hamiltonian $\hat{H}_{{\rm{S}}}$ for finite values of $J,w,m$) is challenging for current experiments, as discussed in section \ref{Sec_ContinuumLimit} below. Therefore, we demonstrate the capabilities of our scheme by studying the coherent real-time dynamics of the creation of particle-antiparticle pairs out of the bare vacuum (the eigenstate of the Schwinger Hamiltonian for $m\rightarrow \infty$) following a quantum quench, i.e., following a rapid change from $m/w=\infty$ to a finite value. 

In the context of particle-antiparticle production, our approach provides the potential to study various interesting quantities in quantum simulation experiments. We consider here two key observables, the vacuum persistence amplitude of the unstable vacuum (section~\ref{SubSec_DecayVacuum}) and the entanglement generated during pair creation (section~\ref{SubSec_Entanglement}). While the former quantity requires only local addressability (which is already needed for implementing the simulation protocol), a measurement of the entanglement entropy is more ambitious due to the increased amount of resources required for reconstructing density matrices using, e.g., quantum state tomography.
This section addresses the phenomenology for the perfect implementation of the proposed simulation scheme. A detailed analysis of the influence of errors and imperfections will be given in section~\ref{Sec_Implementation}.
%
%
\begin{figure}[t]
\centering
\includegraphics[width=\textwidth, angle=0]{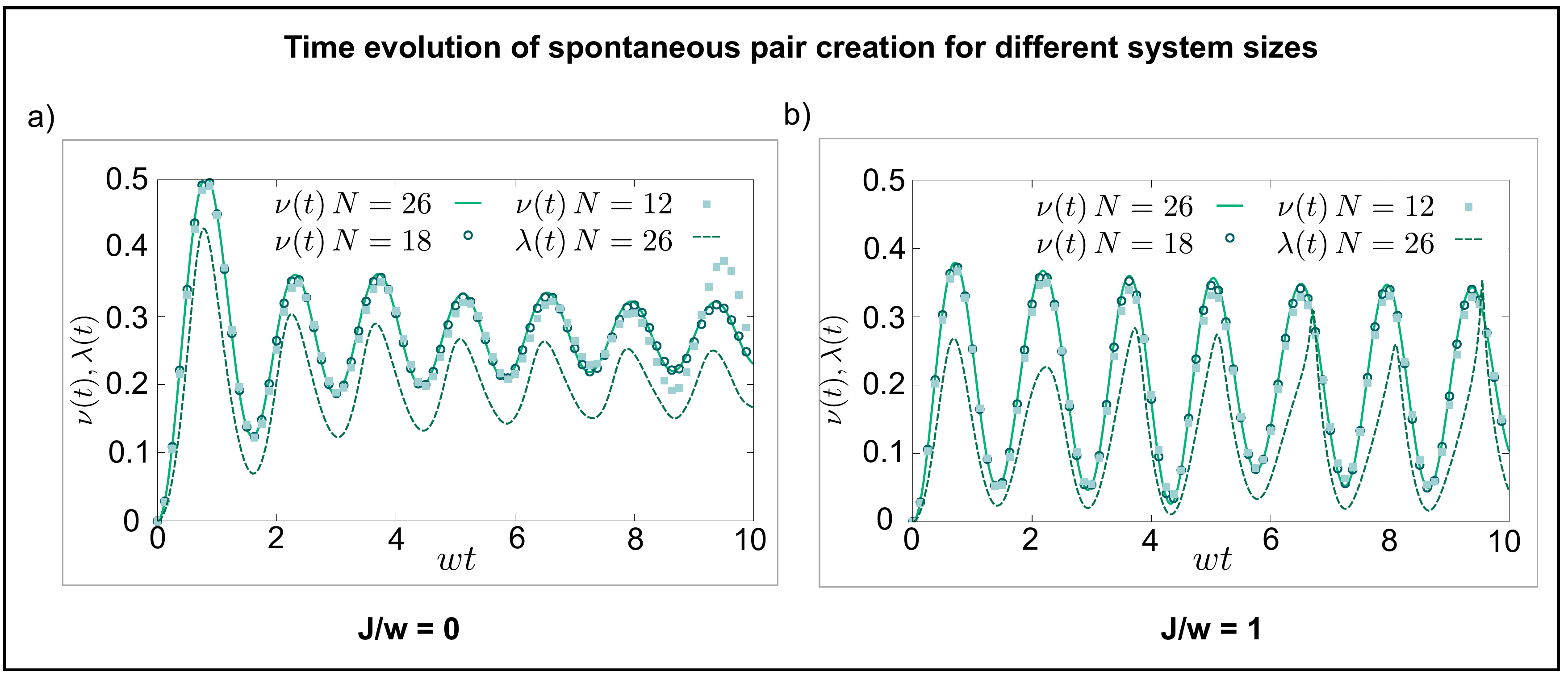}
\caption{\label{Fig_ResultsIdealCase_2} Dynamics of particle production in comparison to the vacuum persistence amplitude. The evolution of the particle number density $\nu(t)$ and the rate function $\lambda(t)$ of the vacuum persistence amplitude, for fermion mass $m/w=1$ and two values of the electric field, (a) $J/w=0$ and (b) $J/w=1$. As $J/w$ is increased, the cost for separating particle-antiparticle pairs rises, resulting in stronger recombination dynamics. This leads to smaller absolute values of $\nu(t)$ and larger oscillations. On the lattice, the one-to-one correspondence between $\nu$ (solid line) and $\lambda$ (dashed), valid in the continuum, is qualitatively retained, even for comparatively small system sizes. The included results for $\nu(t)$ for smaller system sizes (squares: $N=12$, circles: $N=18$) visualize that the dynamics quickly converges with increasing $N$.}
\end{figure}
%
%
\subsection{Decay of the unstable vacuum}\label{SubSec_DecayVacuum}
Since vacuum fluctuations promote the creation of particle-antiparticle pairs, the bare vacuum $|\text{vac}\rangle$ (i.e.\ the state where particles are absent) is unstable. The particle number density $\nu(t)$ created out of the bare vacuum is measured by the observable
\begin{eqnarray*}
\nu(t)=\frac{1}{2N}\sum_{n=1}^N  \langle  (-1)^n \hat\Phi_n^\dag(t) \hat\Phi_n(t) +1\rangle \,
\end{eqnarray*}
with $\langle \dots \rangle =\langle \text{vac} | \dots | \text{vac} \rangle$ denoting the average with respect to the initial vacuum state.  After transforming the fermonic fields to spin operators by a Jordan-Wigner transformation (see section~\ref{SubSec_Encoding}), the particle number density is given by $\nu(t)\!=\!\frac{1}{2N}\sum_{n=1}^N\langle(-1)^n \hat{\sigma}_n^z(t)+1\rangle$ and can be determined through local magnetization measurements (see Fig.~\ref{Fig_ResultsIdealCase_1}(a)). In Fig.~\ref{Fig_ResultsIdealCase_1}(b), the quantum real-time dynamics of $\nu(t)$ is shown, illustrating the instability of the vacuum. 
Initially, particles are produced quickly. After a sufficiently large particle density has been generated, particle-antiparticle recombination becomes favored, inducing a decrease of $\nu(t)$.
This nonequilibrium interplay of regimes with either dominating production or recombination continues over time, leading to an oscillatory behavior of $\nu(t)$ with a slowly decaying envelope. Asymptotically, the system reaches a steady state with a balance between particle production and recombination. As shown in Fig.~\ref{Fig_ResultsIdealCase_1}(a), increasing particle masses lead to a decrease in the particle production because of the increasing energy costs for pair creation. Similarly, with larger values of $J/w$, the higher cost of generating a field string between electrons and positrons reduces the density of generated pairs, see Fig.~\ref{Fig_ResultsIdealCase_2}.\\
\\An important quantity in the context of dynamics in the Schwinger model is the vacuum persistence amplitude~\cite{Schwinger1951}
\begin{eqnarray*}
\mathcal{G}(t)=\langle \text{vac}|e^{-i \hat{H}_{S}t}|\text{vac}\rangle,
\end{eqnarray*}
which measures the deviation from the initial state during the simulated dynamics and quantifies therefore the aforementioned decay of the unstable vacuum. In the continuum limit it has been shown that the decay of the vacuum is directly related to the particle production $\nu(t)= \lambda(t)$ with $\lambda(t) = -N^{-1} \log\left(|\mathcal{G}(t)|^2\right)$~\cite{Schwinger1951}.
In Fig.~\ref{Fig_ResultsIdealCase_2}, we show a comparison between $\lambda(t)$ and $\nu(t)$ for different parameters. While on the lattice their one-to-one relation $\nu(t)= \lambda(t)$ is broken, we find numerically that the similarity between $\lambda(t)$ and $\nu(t)$ nevertheless remains clearly visible. 
Vacuum persistence amplitudes are not only important for spontaneous pair creation. They appear under different names in a variety of contexts in quantum many-body theory and are therefore also of interest in other types of quantum simulations
(for example, in the theory of quantum chaos~\cite{Gorin2006dv} the associated probability $\mathcal{L}(t)=|\mathcal{G}(t)|^2$ is also known as the Loschmidt echo and quantifies the stability of quantum motion~\cite{Peres1984kq}. It plays also a central role in dynamical quantum phase transitions far from equilibrium~\cite{Heyl2013a,Heyl2015dq}. In quantum thermodynamics, the Fourier transform of $\mathcal{G}(t)$ is related to work distribution functions~\cite{Talkner2007aw}, which are the basic objects appearing in nonequilibrium fluctuation theorems~\cite{Campisi2011fk} such as the Jarzynski equality~\cite{Jarzynski1997zl}).
%
%
\begin{figure}[t]
\centering
\includegraphics[width=0.75\textwidth, angle=0]{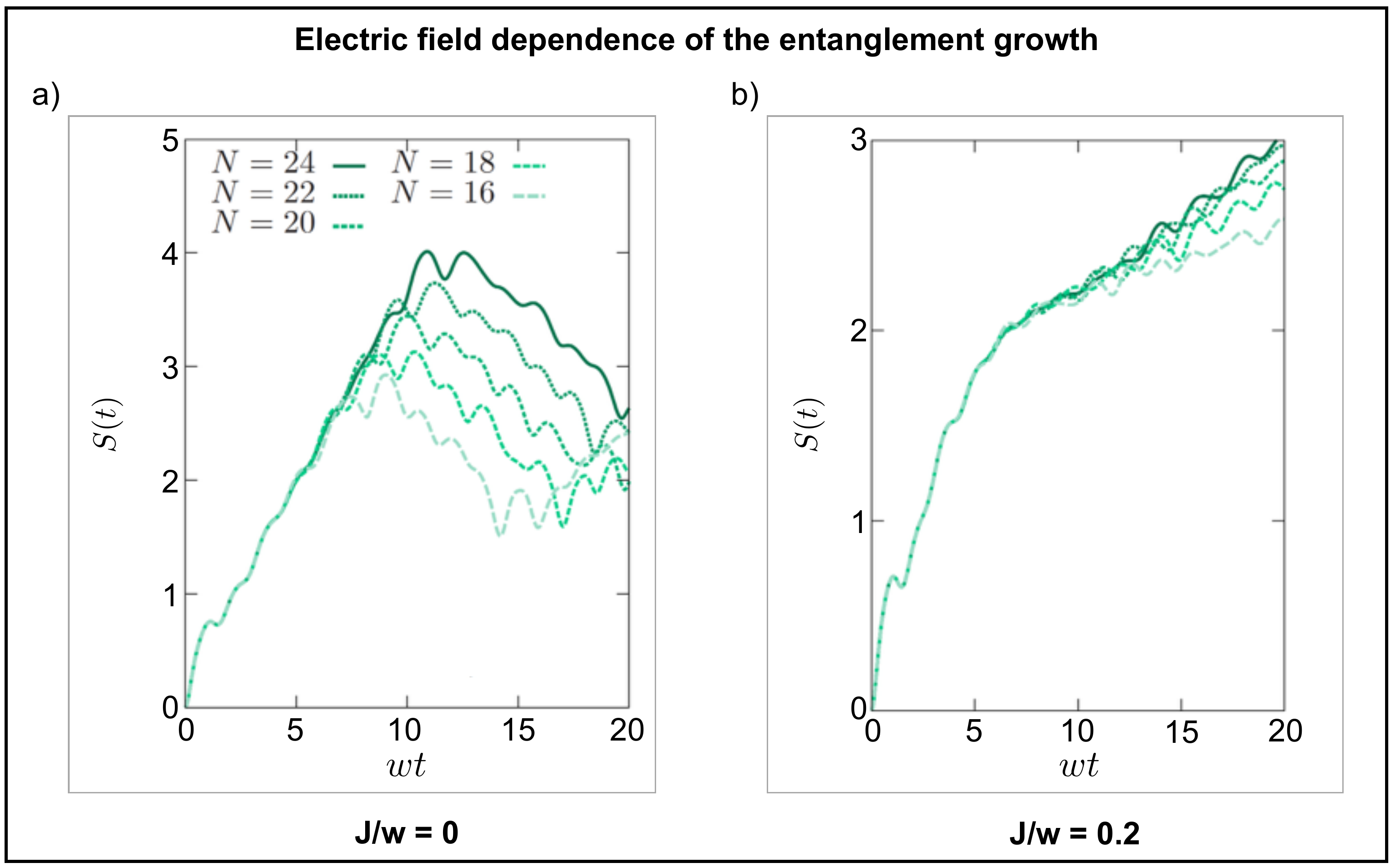}
\caption{\label{Fig_ResultsIdealCase_3} Time Evolution of the half chain entanglement entropy $S(t)$ for different system sizes $N=\{16, 18, 20, 22, 24\}$ and for
two different electric field strengths, (a) $J/w=0$ and (b) $J/w=0.2$ (notice the different $y$-axis scalings for both panels), with the fermion mass set to $m/w=1$. (a) At vanishing field energies, particles spread ballistically, leading to a linear increase of entanglement over time. This increase is cut off by finite system sizes on a time scale $wt \propto N/2$.
(b) The energy cost for generating particle-antiparticle pairs, as well as for separating them, increases with $J/w$, thus suppressing the amount of entanglement that is generated. The ballistic linear increase can only bee seen over time scales on the order of $t_J = J^{-1}$.
}
\end{figure}
%
%
\subsection{Entanglement dynamics in spontaneous particle production}\label{SubSec_Entanglement}
A further important quantity for the theoretical characterisation of quantum many-body dynamics is given by the entanglement entropy~\cite{Lamacraft2012}. In the following, we study the real-time entanglement production during pair creation. We focus on the entanglement between two contiguous blocks in the spin system, which is equivalent to the entanglement between the respective blocks in the original fermionic description including the gauge degrees of freedom (see~\ref{App_Entanglement} for details). Indeed, when the system is encoded, the Hilbert space takes a factorized form (while in the initial formulation, it is not factorized due to the presence of Gauss's law); this form recovers the construction proposed in~\cite{Casini2014}. Therefore, it is possible to measure the entanglement in the original model directly in a pure spin system, which has the advantage that the partition into subsystems is always gauge invariant thereby avoiding known difficulties with tracing out parts of the system~\cite{Casini2014}.

In the following, we consider the real-time dynamics of the half chain entanglement entropy,
\begin{equation}
	S(t) = -\mathrm{tr}_A \left\{\rho_A(t) \log\left[ \rho_A(t) \right]\right\}
\end{equation}
with $\rho_A=\tr_B\left\{ \rho(t)\right\}$ denoting the reduced density matrix of the first $N/2$ lattice sites, obtained by tracing out the remaining system $B$. $S(t)$ quantifies the entanglement between the left and the right half of the system, which is generated by the creation of particle-antiparticle pairs that are distributed across the two parts. As already shown in Fig.~\ref{Fig_ResultsIdealCase_1}(c), the generation of entanglement decreases with increasing mass $m$, since particle creation becomes energetically costly for a large fermion mass.
The dependence of the entanglement production in the electric field energy $J$ is particularly interesting. The effective coupling between particles and antiparticles, mediated by the gauge bosons, increases with their relative spacing such that it becomes energetically unfavourable to separate particle-antiparticle pairs over large distances. This constrains the dynamics by reducing the number of particle-antiparticle pairs that can be shared between the left and the right half of the system. Accordingly, this leads to a reduction of entanglement for increasing $J$. In Fig.~\ref{Fig_ResultsIdealCase_3}, we show the time evolution of the half chain entanglement entropy $S(t)$ for two different electric field strengths $J/w=0$ and $J/w=0.2$ for varying system sizes.
For the case of free particles, i.e., without coupling to the gauge fields ($J=0$), the entanglement entropy exhibits a linear growth in time, characteristic for free fermionic theories. 
In the thermodynamic limit $N\rightarrow\infty$, the linear growth would continue for all times, but for finite system sizes $N<\infty$ it is cut off  on a time scale $wt \propto N/2$. 
If $J\neq 0$, the entanglement growth follows initially approximately the free case up to a time scale $t_J = J^{-1}$, beyond which the entanglement production is substantially slowed down. As the effective potential experienced by a particle-antiparticle pair increases linearly in the distance for nonzero $J$, the electric field suppresses the separation of spontaneously generated pairs over large distances. This, in consequence, reduces the amount of entanglement that can be produced. As these examples show, the quantum simulation of the Schwinger model allows for the observation of an intricate interplay between different parameter regimes, which can be studied in quantities not accessible to conventional experiments.

\section{Imperfections of the scheme and implementation in trapped ions}\label{Sec_Implementation}
In the following, we describe how the proposed quantum simulation protocol can be implemented in a system of trapped ions (see section~\ref{SubSec_Experiment}) and discuss the effects of imperfections. There are two types of errors: (i) those that are inherent to the scheme and (ii) those that are due to experimental imperfections. The former, discussed in section~\ref{SubSec_TrotterErrors}, arise since our digital quantum simulation protocol realizes the desired dynamics only in a time-averaged manner, which leads to a discretization error~\cite{Trotter59,Lloyd96,Sanders11,Sanders12} also known as Trotter error. The latter depend on the concrete physical implementation. In section~\ref{SubSec_OtherExpErrors} and section~\ref{SubSec_Decoupling}, we discuss the sensitivity of the simulations to experimental imperfections by considering the trapped-ion implementation that has been realized in~\cite{ExpPaper}. We identify dominant errors and show that the phenomena of interest are robust against the main sources of imperfections that generically occur in this type of setup.
\subsection{Implementation of the simulation scheme using trapped ions}\label{SubSec_Experiment}
In trapped-ion setups, spin degrees of freedom are obtained by restricting the dynamics to two (meta-) stable Zeeman levels of the internal electronic level structure of the
ions~\cite{Blatt2012,Schindler2013}. The use of focused laser beams acting on these levels allows one to realize single-qubit operations by inducing individually addressed AC-Stark shifts ($\hat{H}^{n}_{\text{\tiny{AC}}}\sim \hat{\sigma}_n^z$) and Rabi flops ($\hat{H}^{n}_{\text{\tiny{RF}}(\theta)}\sim \cos(\theta) \hat{\sigma}_n^x+\sin(\theta)\hat{\sigma}_n^y$). Spin-spin interactions are realized using a global laser field coupling the internal levels to external vibrational degrees of freedom. Here, we assume an
implementation based on the so-called M\o lmer-S\o rensen
interaction~\cite{MSgates,Milburn99,Solano99,Leibfried2003} that is mediated by the
motional center of mass mode~\cite{Blatt2012} and provides an infinite
range, all-to-all two-body coupling
$\hat{H_0}=J_0\sum_{n,l}\hat{\sigma}_{n}^x\hat{\sigma}_l^x$, as
assumed in section~\ref{SubSec_Protocol}.
This effective description of the spin-spin interaction neglects the motional degrees of freedom of
the chain of ions, which is valid if the spin-motion coupling after a
gate operation is negligible and the duration of the interaction is
much larger than the period of the harmonic motion of the ions in the
trap. This condition is well fulfilled in the considered experimental setting~\cite{Blatt2012,Schindler2013}, as the period of the
harmonic motion is typically less than 1$\mu s$ and the gate duration is longer
than 50$\mu s$.

Our simulation protocol requires the decoupling of individual spins from the infinite-range coupling $\hat{H_0}=J_0\sum_{n>l}\hat{\sigma}_{n}^x\hat{\sigma}_l^x$ (see Fig.~\ref{Fig_SimulationProtocol}(c),(d)), which can be achieved in different ways. 
For example, one may strongly detune individual ions from the laser fields that induce the M\o lmer-S\o rensen interaction by applying strong addressed AC Stark shifts \cite{Smith2015,InteractingQuasiParticles}. 
Alternatively, one can split the ion crystal into multiple chains during a single
experiment, such that only the ions that take part in the long-range
interaction form a connected crystal that interacts with the laser light. This
flexible scheme requires the use of micrometer-scale ion traps which increases
the experimental complexity considerably~\cite{Walther2012,Bowler2012}.
An alternative that can be implemented in a macroscopic trap, and which has been employed in \cite{ExpPaper}, consists in the transfer of the population of the
idling ions to additional electronic substates that are
off-resonant with respect to the laser light inducing the gate operations~\cite{Schindler2013}. This decoupling (recoupling) procedure will be referred to as hiding
(unhiding) below.
%
\begin{figure}[t]
\centering
\includegraphics[width=\columnwidth, angle=0]{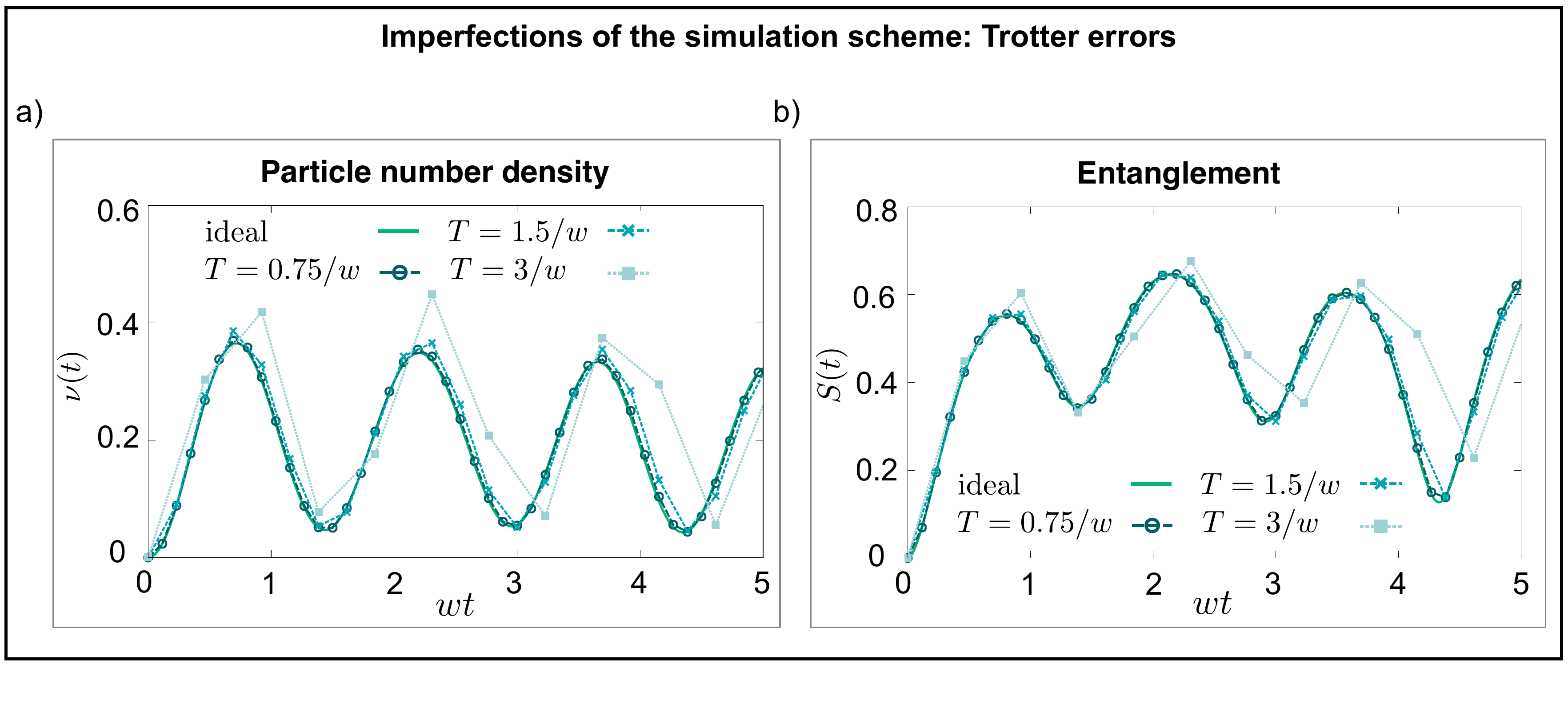}
\caption{\label{Fig_5_1} Quantum simulation of spontaneous pair creation for finite Trotter step size $T$  (see Fig.~\ref{Fig_SimulationProtocol}a). We show the evolution of the particle number density $\nu(t)$ in panel (a) and the half chain entropy $S(t)$ in panel (b) for $m=J=w$ and $N=10$, with step sizes $T=0.75/w$ (circles) $T=1.5/w$ (crosses), $T=3/w$ (squares) and for the ideal case (dark green solid line). As the step size decreases, the results converge fast towards the ideal case. The evolution time $wt=5$ corresponds to $t=16$ms if a M\o lmer-S\o rensen coupling strength $J_0=4$kHz is assumed.}
\end{figure}
%
\subsection{Discretization errors}\label{SubSec_TrotterErrors}
The digital quantum simulation scheme introduced in section~\ref{SubSec_Protocol}, allows one to realize the time evolution under the Hamiltonian $\hat{H}_{\rm{S}}$ by means of a stroboscopic sequence, which consists of the cyclic application of Hamiltonians that can be experimentally realized $\hat{H}_1, \hat{H}_2,...\hat{H}_n$ (see Fig.~\ref{Fig_SimulationProtocol}). The Trotter error, i.e.\ the difference between the desired time evolution $U_{{\rm{S}}}=e^{-i\hat{H}_{{\rm{S}}} t}$ and the evolution realized by the stroboscopic sequence $U_{{\rm{sim}}}=\left(e^{-i\hat{H}_1t/n}...e^{-i\hat{H}_nt/n}\right)^n$ is bounded by~\cite{Lloyd96}
\begin{eqnarray*}
U_{{\rm{S}}}-U_{{\rm{sim}}}=\frac{t^2}{2n}\sum_{i,j}[\hat{H}_i,\hat{H}_j]+\epsilon,
\end{eqnarray*}
where $\epsilon$ represents higher-order
terms\footnote{$\epsilon=\sum_{k=3}^{\infty}E(k)$, where
  $||E(k)||_{{\rm{max}}}\le n||Ht/n||^k_{{\rm{max}}}/k!$, and
  $||O||_{{\rm{max}}}$ is the maximum expectation value of the
  operator $O$ (see~\cite{Lloyd96}).}. Hence, these errors are
controllable and the accuracy of the Trotter decomposition can, in
principle, be increased to any desired precision by increasing the
number of time steps $n$. 
However, the implementation of the proposed scheme in trapped ions poses limits on the minimum length of the step size that can be used. In the presence of decoherence, this leads to practical limitations on the accuracy with which the dynamics can be realized. More specifically, in order to suppress undesired spin-motion coupling terms during the applied M\o lmer-S\o rensen gates (see section~\ref{SubSec_Experiment}), we require a minimal length $\Delta t_{\text{\tiny{min}}}$ of the basic time windows that are used, such that $\omega_{{\rm{trap}}}\Delta t_{\text{\tiny{min}}}\gg1$. In the following, we consider typical experimental values, where $\omega_{{\rm{trap}}}$ takes values on the order of MHz~\cite{Blatt2012,Schindler2013} and evaluate the Trotter error numerically for $\Delta t_I, \Delta t_{II} \gg\omega^{-1}_{{\rm{trap}}}$ (compare Fig.~\ref{Fig_SimulationProtocol}(c),(d)). Local operations can be performed about an order of magnitude faster than entangling operations~\cite{Schindler2013}, and thus, one-qubit rotations are assumed to be instantaneous in our model.  As Fig.~\ref{Fig_5_1} and Fig.~\ref{Fig_5_2} demonstrate, the Trotter error can be made sufficiently small to obtain a good resolution of the relevant features (see also experimental results in~\cite{ExpPaper}). 
Thus, this error intrinsic to digital quantum simulation is well controlled and not a limiting factor.
\subsection{Main experimental errors in trapped-ion implementations}\label{SubSec_OtherExpErrors}
%
\begin{figure}[t]
\centering
\includegraphics[width=\columnwidth, angle=0]{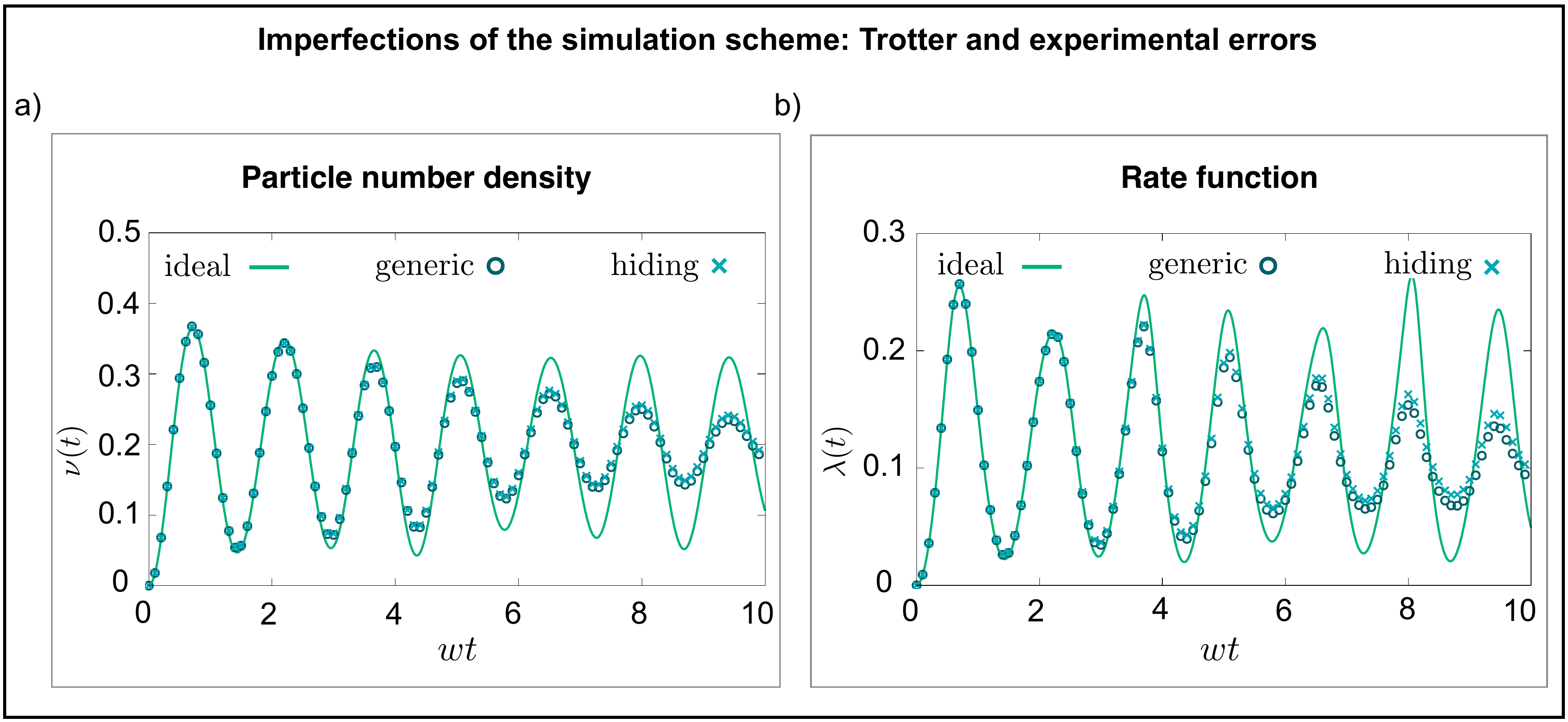}
\caption{\label{Fig_5_2} Quantum simulation of spontaneous pair creation including experimental imperfections. Panel (a) shows the evolution of the particle number density $\nu(t)$ for $N=10$ as a function of the dimensionless time $wt$ for $m=J=w$ (see Eqs.~(\ref{Eq_H_E-field})-(\ref{Eq_H_LocalTerms})) for a Trotter step size $T=1.3/w$. Circles correspond to results including fluctuations of the M\o mer-S\o rensen coupling strength $J_0$ of magnitude $\delta J_0\in[-0.05,0.05 ]J_0$ and collective dephasing of order $\delta\omega\in[-0.025,0.025 ]J_0$ as explained in section~\ref{SubSec_OtherExpErrors}. Crosses represent results that take the different dephasing strength of ions that are transferred to hiding levels into account (see section~\ref{SubSec_OtherExpErrors}). We consider here the implementation in~\cite{ExpPaper} and assume a dephasing strength of $\delta\omega'=1.5\delta\omega$ for ``hidden" ions. Panel (b) shows the rate function $\lambda(t)$ as is defined in section~\ref{SubSec_DecayVacuum} for the same set of parameters.}
\end{figure}
In this subsection, we discuss generic errors affecting the gate fidelity of the digital quantum simulation in trapped-ion systems. 
In the considered type of experiment~\cite{Blatt2012,Schindler2013}, the two main error sources
impairing the required gate operations are (i) fluctuating coupling
strengths of the M\o lmer-S\o rensen interaction and (ii) collective dephasing (see below). 
The first imperfection, fluctuations in the coupling strength $J=(J_0+\delta J)$, is
mainly due to intensity and beam pointing fluctuations of the laser
beam, resulting in slightly modified interactions $\hat{H_0}=(J_0+\delta J)\sum_{n,l}\hat{\sigma}_{n}^x\hat{\sigma}_l^x$. The laser intensity fluctuations are
slow on the time scale of a single simulation experiment, such that
 $\delta J$ can be considered to be constant within one experimental run.
Consequently, the experiment effectively realizes an ensemble of
coherent Hamiltonian evolutions with fluctuating interaction
strength. The resulting expectation values are averages over
trajectories corresponding to different values of $\delta J$. 

The second major source of imperfections, dephasing between the two Zeeman levels that encode the spin states, is induced by fluctuating
magnetic fields. Since the ions are typically only micrometers apart, all spins experience
approximately the same field fluctuations, and the effect can be described in
terms of Hamiltonian perturbations of the type
$\hat{H}_{{\rm{deph}}}=\delta{\omega}\sum_n\hat{\sigma}_n^z$ with
randomly varying coefficients $\delta{\omega}$. Again, on the time
scale of a single experiment, $\delta \omega$ can be assumed to be
time-independent such that we can take these experimental
imperfections into account by averaging over an ensemble of
trajectories for different values of $\delta \omega$. It is interesting to note
that the considered ideal time evolutions (starting from the bare
vacuum state shown in Fig.~\ref{Fig_ResultsIdealCase_1}a) take place
in the zero magnetisation subspace, which forms a
decoherence-free-subspace with respect to collective dephasing\footnote{In the
absence of spin flips or undesired state transfers, the quantum states
after a full time step are therefore invariant under this type of noise. However, as explained in section~\ref{SubSec_Protocol}, the active qubits
undergo a rotation $U_y$ (see Fig.~\ref{Fig_SimulationProtocol}c) that changes the reference frame, such that the noise after
this rotation is effectively given by
$U_y\hat{H}_{{\rm{deph}}}U_y^{\dag}=\delta{\omega}\sum_n\hat{\sigma}_n^x$, an
interaction that induces spin flips.}.

To estimate the influence of these generic imperfections, we perform a
numerical simulation of the expected time evolution for the particle
number density $\nu(t)$ and the rate function $\lambda(t)$. To this end, 
we draw $\delta J$ and $\delta \omega$ randomly from uniform
distributions. Fig.~\ref{Fig_5_2} shows calculated data for a chain of ten ions
with $J/w=m/w=1$.  The solid line corresponds to the ideal case, while
the circles represent the expected results taking the discussed
experimental imperfections into account. Here, we assumed fluctuations
of the M\o lmer-S\o rensen coupling rate of $5\%$
($\delta J \in [-0.05,+0.05]J_0$) and a collective dephasing rate of
$2.5\%$ ($\delta \omega \in[-0.025,+0.025]J_0$).  Assuming a M\o
lmer-S\o rensen coupling rate of $J_0=4$kHz, the circles correspond to
Trotter steps where the elementary time window $T$ (see
Fig.~\ref{Fig_SimulationProtocol}) is $0.325$ms long. 
These fluctuations lead to a damping of the amplitude of the oscillations, though the frequency is retained. 
As these results show, the qualitative agreement including these errors remains satisfactory through realistic evolution times. 

Additionally, if part of the qubit register is in the hiding states
during a many-body interaction (see section~\ref{SubSec_Experiment}), different phase shifts should be taken
into account for ions that participate in the M\o lmer-S\o rensen
interaction and for those that are transferred to hiding levels. 
Fig.~\ref{Fig_5_2} contains a numerical simulation including this effect (crosses), 
where we consider the internal levels used in~\cite{ExpPaper}. These results have been obtained by adding terms
$\hat{H}_{{\rm{deph}}}=\delta{\omega}\sum_n\hat{\sigma}_n^z+\delta{\omega}'\sum_l\hat{\sigma}_l^z$
for each time window, where the first (second) sum includes ions that
occupy regular (hiding) states. As Fig.~\ref{Fig_5_2} shows, this
leads only to minor corrections. 
The error model above captures the dominating sources of
imperfections in the considered setting. The effect of imperfect local
operations on non-hidden qubits is negligible compared to the errors
discussed above, since they can be performed with a much higher
accuracy than multi-qubit entangling gate operations. In particular, if only a
finite set of local operations is required, fidelities larger than
$F_{{\rm{local}}}=0.99$ can be reached~\cite{Schindler2013}.

\subsection{Error detection techniques}\label{SubSec_Decoupling}

The use of hiding/unhiding techniques (see section~\ref{SubSec_Experiment}) generates another source of imperfections in the simulated time evolution. 
If the applied hiding pulses fail to transfer a quantum state from the computational basis states to the corresponding hiding states\footnote{If the pulse area of a pulse deviates from the target value, an imperfect state transfer is performed  $\ket{\uparrow}\rightarrow\cos(\frac{\pi}{2}+\delta)\ket{\uparrow}+\sin(\frac{\pi}{2}+\delta)\ket{h_{\uparrow}}$. If the qubit is measured after a series of imperfect pules, the resulting erroneous state can be associated with a failure probability $p=\sin(\delta)^2$ for each pulse.}
$\ket{\uparrow}\rightarrow \ket{h_{\uparrow}}$,
$\ket{\downarrow}\rightarrow \ket{h_{\downarrow}}$, or unhiding pulses fail to transfer the quantum states back, the many-body
operation will not act on the ions as intended and thus induces many-body errors that are difficult to correct for. 
However, as we describe in the following, postselection techniques can be employed to detect and filter out this type of error without affecting the desired unitary evolution.
Such a postselection scheme can for example be realized as follows. For each step in the protocol that involves a M\o lmer-S\o rensen gate acting on a subset of ions, suitable hiding operations are performed, followed by the action of the quantum gate, and the corresponding unhiding operations. Following these steps, a measurement of the population residing in the hiding levels is performed. 
If population is detected in the hiding levels, the experimental run is discarded. This way, only events that involve a failure of both the hiding and the unhiding pulse on a single ion (in a single hiding-unhiding step)
remain undetected, which, however is strongly suppressed~\footnote{The probability for such an event is given by $P_{1}=p^2 N$ for each hiding-unhiding step, where $p$ is the  
probability for a hiding or unhiding pulse to fail, assumed to be small. The probability for such an undetectable error to occur during a time window $T$ scales therefore with $N^3p^2$.}.
Apart from a controllable residual error that can be suppressed to any
desired accuracy, decoupling errors lead therefore only to a reduction
of the rate at which simulation data can be acquired, but do not impair
the desired unitary evolution.

In~\cite{ExpPaper}, an alternative postselection scheme has been applied to ensure that the final state fulfills Gauss' law. This involves measurements of the total magnetisation
$\hat{M}= \sum_n\hat{\sigma}_n^z$ at the end of simulated time
evolution. Due to charge conservation in the Schwinger model,
$\hat{M}$ is a conserved quantity under the dynamics induced by
$\hat{H}_{{\rm{S}}}$. The simulations
performed in~\cite{ExpPaper} have been carried out in the
zero-charge subspace, starting from the initial state
$\ket{\uparrow\downarrow\uparrow\downarrow...}$. Using the employed decoupling and measurement techniques (see~\cite{ExpPaper,Schindler2013}), single hiding/unhiding errors can lead to measurement results that correspond to a nonzero magnetisation of the spin system. This subset of errors could therefore be detected and filtered out by postselection.

We remark that digital quantum simulation schemes allow one in principle to use quantum error correction schemes to ensure that the desired sequence of gates is carried out with high fidelity. The postselection techniques discussed here provide an alternative route that is easier to realize experimentally and allows one to filter out the dominant errors.
%
%
\section{Continuum limit}\label{Sec_ContinuumLimit}

%
\begin{figure}[t]
\centering
\includegraphics[width=0.8\columnwidth, angle=0]{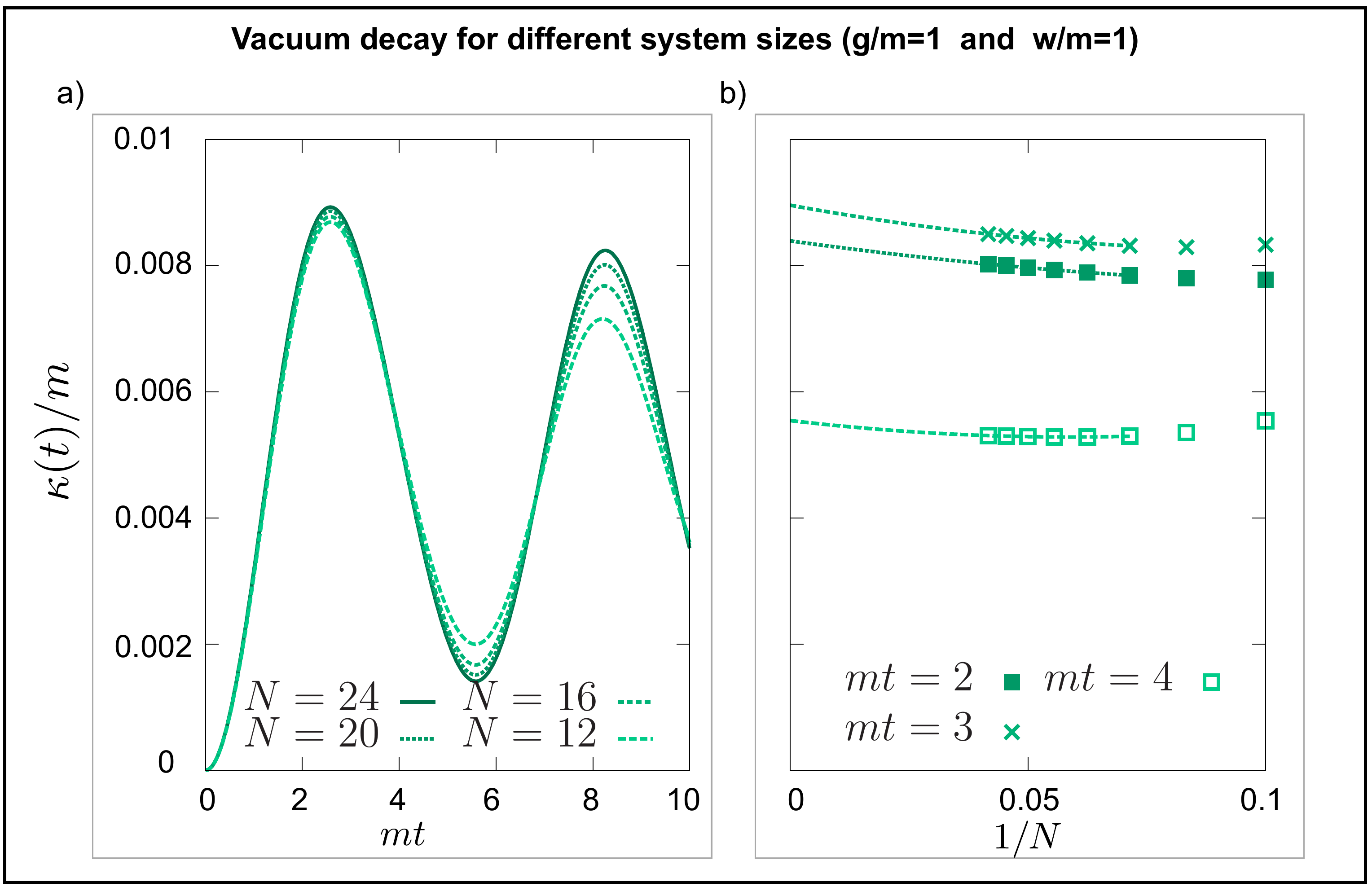}
\caption{\label{Fig_6_1} Loschmidt echo rate function $\kappa(t)$ per unit length for $g/m=1$ and $m/w=1$. Panel (a) compares the dynamics of $\kappa(t)$ for different system sizes. Panel (b) illustrates the extrapolation of the data to the thermodynamic limit for a few selected time points. Lines are fits including corrections of orders $1/N$ and $1/N^2$. The lines only extend over those data points that have been included in the fit.}
\end{figure}
%
%
\begin{figure}[t]
\centering
\includegraphics[width=0.8\columnwidth, angle=0]{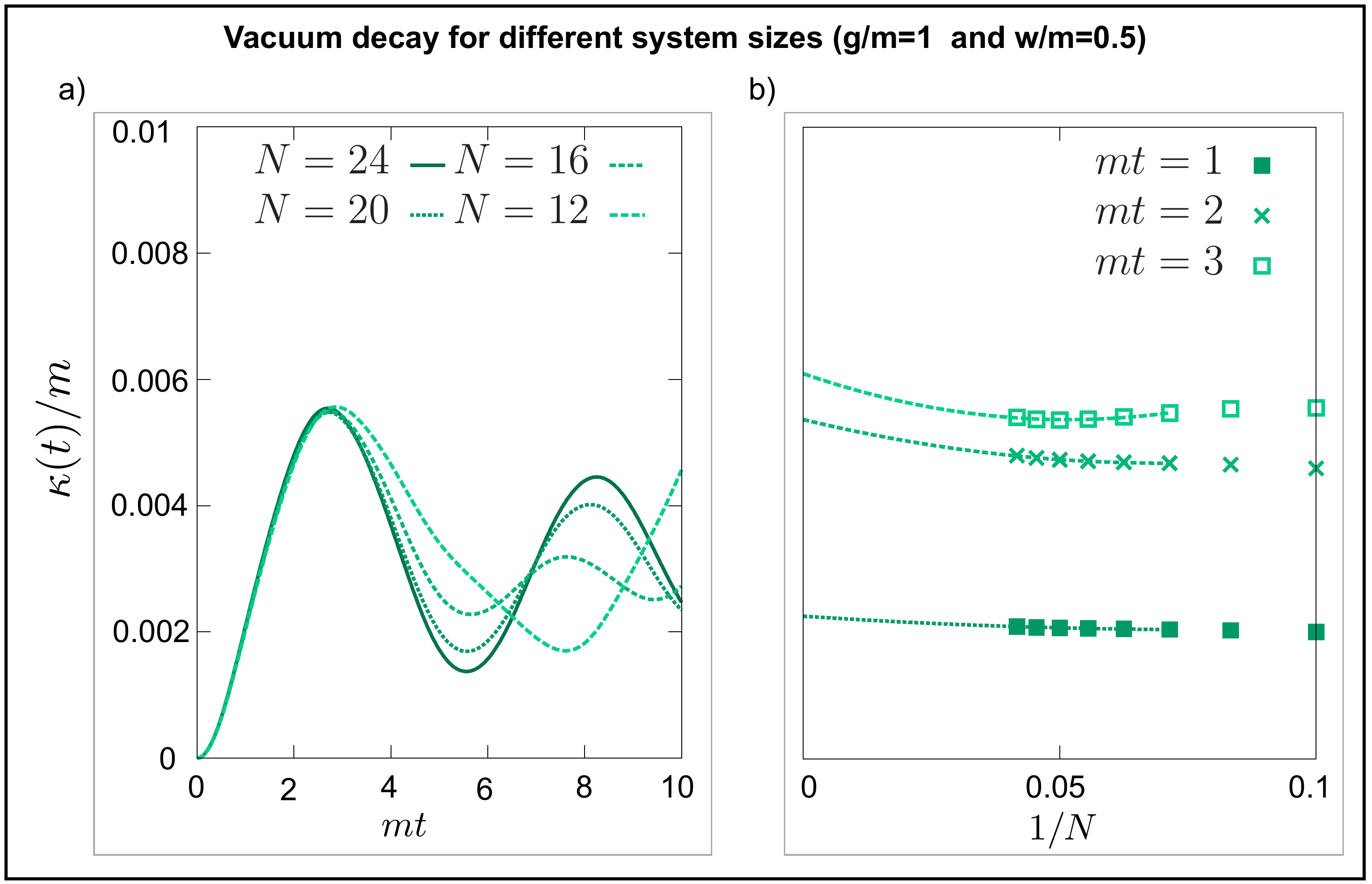}
\caption{\label{Fig_6_2} Loschmidt echo rate function $\kappa(t)$ per unit length for $g/m=1$ and $m/w=0.5$ (halved lattice spacing compared to Fig.~\ref{Fig_6_1}). Panel (a) compares the dynamics of $\kappa(t)$ for different system sizes, showing that finite-size effects become more severe with decreasing lattice constant. Panel (b) presents extrapolations of the data for a few selected time points, where lines are fits including corrections of order $1/N$ and $1/N^2$.}
\end{figure}
%
In the previous sections, we have shown how to simulate the Schwinger model within its lattice formulation. 
As we will discuss in the following, the ultimate goal, the extrapolation from the lattice to the original continuum field theory, can in principle be reached, though it remains a challenging prospect.  
The continuum limit $a\to 0$, where $a$ is the lattice spacing, involves a rescaling~\cite{Encoding} of the couplings in the lattice Schwinger Hamiltonian $\hat H_\mathrm{lat}$ given in Eq.~(\ref{Eq_LatticeHamiltonian}), 
\begin{equation}
  w = \frac{1}{2a}, \quad J=\frac{g^2}{2} a\,, 
\end{equation}
while the rest mass $m$ is independent of $a$. To correctly reproduce the original Schwinger model, the thermodynamic limit $N\to\infty$ has to be taken  before taking the limit $a\to 0$. Practically, both for theoretical or experimental data, this specific order of limits can be implemented by first fixing a lattice constant $a$ and then extrapolating the data to the thermodynamic limit. In this way, data can be obtained successively for decreasing lattice spacings $a$, which then finally may be extrapolated to $a\to 0$.

Taking the continuum limit requires an initial state that correctly reproduces the long-wavelength properties of the continuum theory. 
While the bare vacuum state chosen in the previous sections is valuable to study the dynamics of the lattice Schwinger model, it exhibits a spatial modulation at the level of the lattice spacing far beyond the long-wavelength limit. 
In the following, we will therefore choose a different initial state that lies in the long-wavelength sector while still being accessible experimentally. 
More specifically, we start from the ground state for the parameters $m>0$, $w>0$ and $g=0$, and then quench to $g/m>0$. 
The system may be initialized in the desired state via adiabatic state preparation. For that purpose the system is initialized in a N\'eel-type state $\ket{\uparrow\downarrow\uparrow\downarrow\dots}$, which can be prepared easily as well as with high fidelity~\cite{Schindler2013} and which is also the ground state of the Hamiltonian $\hat{H}_{\rm{m}}=(m/2)\sum_n (-1)^i\hat{\sigma}_n^z$. Afterwards, the Hamiltonian is adiabatically transformed from $\hat{H}_{\rm{m}}$ to  $\hat{H}=\hat{H}_{\pm}+(m/2)\sum_n (-1)^i \hat{\sigma}_n^z$ by engineering a time-dependent Hamiltonian $\hat{H}(t) = \hat{H}_m + f(t) \hat{H}_{\pm}$ with $f(0)=0$ and $f(t')=1$ where $t'$ denotes the time of the end of the process. For a sufficiently slow transformation the state always remains in the ground state manifold of the instantaneous Hamiltonian according to the adiabatic theorem~\cite{AdiabatenSatz}. The error in following the adiabatic passage is mainly set by the minimal gap of the Hamiltonian $\hat{H}(t)$ during the sweep. Importantly, the considered transformation will not encounter a quantum phase transition and therefore no gap closing. As a consequence, the considered adiabatic state preparation is very efficient. Instead of conventional adiabatic state preparation, which is based on a continuous deformation of the Hamiltonian over time, this procedure can also be realized in a digital quantum simulation device, as demonstrated recently with superconducting qubits~\cite{Barends2016}. When choosing the digital approach also for the preparation of the initial state, the overall runtime of the quantum simulation, i.e., the total number of gate operations, increases respectively. Fortunately, the number of gate operations per Trotter step for the preparation is $N$, which is smaller than for the digital simulation for the full Schwinger model which is $2N-2$.

In the following, we will consider the continuum limit exemplarily for the vacuum persistence probability $\mathcal{L}(t) = |\mathcal{G}(t)|^2$ with $\mathcal{G}(t)=\langle \psi_0 | e^{-i\hat H_s t} | \psi_0 \rangle$. In the limit $a\to0$, the rate function
\begin{equation}
      \kappa(t) = -\frac{1}{L} \log\big[ \mathcal{L}(t) \big] 
\end{equation}
has a well-defined limit, where $L=aN$ is the total length of the system. In Fig.~\ref{Fig_6_1}, we show the time evolution of $\kappa(t)$ that is induced by switching-on of the electric-field coupling for different system sizes. The chosen parameters correspond to $m/g=1$ and $m/w = 1$. As mentioned before, establishing the continuum limit requires a successive extrapolation to the thermodynamic limit for decreasing values of the lattice spacing $a$. In Fig.~\ref{Fig_6_1}(b), we perform this extrapolation for a few specific time points. For the system sizes available numerically, we find that the leading order $1/N$ correction is not always sufficient for this purpose, and that also the next-to-leading order contribution $1/N^2$ has to be considered to yield good fits for all data points. 
As next step towards the continuum limit, we decrease the lattice spacing by a factor of $2$, yielding $m/w = 0.5$. The corresponding data is shown in Fig.~\ref{Fig_6_2}. Compared to the previous case, finite-size effects become stronger, especially for larger times. For $mt \lesssim 3$, the extrapolation to $N\to \infty$ can be performed, as shown in the right panel of Fig.~\ref{Fig_6_2}. Nevertheless, as becomes clear from these data, a fully reliable limit $N\to\infty$ requires even larger systems, which is beyond the capabilities of the utilized numerics.

These results show that it is in principle possible to obtain the continuum limit. The main long-term challenges in this context are the preparation of the initial state as well as the need for large system sizes, which becomes more severe for decreasing lattice constant $a$.
\section{Conclusions and outlook }\label{Sec_Conclusions}
In this article we described and analysed a protocol for the digital quantum simulation of the Schwinger model, a U(1)-Wilson lattice gauge theory, that has recently been proposed and experimentally demonstrated in~\cite{ExpPaper}. Our scheme is based on encoding the gauge degrees of freedom in a spin chain with long-range interactions, and thus exploits a strategy that has been used previously in numerical calculations. 
We have shown how the encoded Hamiltonian with its complex long-range interactions can be realized using resources available in state-of-the-art digital quantum simulators, requiring only addressable single-qubit manipulations and one type of two-qubit gate. By construction, the protocol retains exact gauge invariance of the dynamics. Furthermore, it realizes a quantum simulation of $2N-1$ degrees of freedom ($N$ matter fields and $N-1$ gauge fields) using only $N$ physical qubits. As we have shown, the number of gate operations scales only linearly with system size, thus allowing for an efficient scaling of such experiments to larger chain lengths. We have performed a careful analysis of potential error sources.  
Through numerical finite-size scaling analyses, we found that already mesoscopic quantum simulators can reliably reproduce the real-time dynamics relevant for larger systems.

Such quantum simulations open exciting prospects, as they are able to extract observables that are not accessible to conventional experiments, such as the vacuum persistence amplitude and entanglement entropy. 
Thus, we can expect to obtain new insights, e.g., into the propagation of entanglement in out-of-equlibrium dynamics. In systems with short-range interactions, the Lieb--Robinson theorem restricts correlations to spread only within sound cones~\cite{Lieb1972}. This restriction does not apply for long-range interactions~\cite{Hastings2006a}, and different dynamical regimes can be observed, for example if interactions follow a power law~\cite{Hauke2013a,Richerme2014,Jurcevic2014}. It will be interesting to study how the possible dynamics is affected by the exotic type of long-range interactions that govern gauge theories. Vacuum persistence amplitudes appear in a wide range of contexts in quantum many-body theory ranging from quantum chaos~\cite{Gorin2006dv} to dynamical quantum phase transitions far from equilibrium~\cite{Heyl2013a}. Therefore it is an interesting question to which extent further insights into the dynamics of lattice gauge theories can be gained by exploring the connections to these other concepts.
Finally, in view of the long-term goal to simulate lattice gauge theories using controlled quantum systems, it will be very valuable to explore avenues to generalize the proposed scheme to two spatial dimensions or ladder geometries, and to non-Abelian gauge theories.\\
\\
\\\textit{Acknowledgements.} \\
We thank D. Banerjee, E. Rico and U.-J. Wiese for useful discussions. We acknowledge support by the Austrian Science Fund (FWF), through the SFB FoQuS (FWF Project No. F4002-N16 and No. F4016-N23), by the European Commision via the integrated project SIQS and the ERC synergy grant UQUAM, by the Deutsche Akademie der Naturforscher Leopoldina (Grant No. LPDS 2013-07 and No. LPDR 2015-01), as well as the Institut f\"ur Quanteninformation GmbH. E.A.M. is a recipient of a DOC fellowship from the Austrian Academy of Sciences. P.S. was supported by the Austrian Science Foundation (FWF) Erwin Schr\"odinger Stipendium 3600-N27. This research was funded by the Office of the Director of National Intelligence (ODNI), Intelligence Advanced Reasearch Projects Activity (IARPA), through the Army Research Office grant W911NF-10-1-0284. All statements of fact, opinion or conclusions contained herein are those of the authors and should not be construed as representing the official views or policies of IARPA, the ODNI, or the U.S. Government.

\appendix

\section{Entanglement in the encoded Schwinger model}\label{App_Entanglement}
We consider the Schwinger model in its encoded version, as described in section~\ref{SubSec_Encoding}, and are interested in evaluating the entanglement between two adjacent regions in space. As explained below, the entanglement between two contiguous blocks of the reduced spin system (i.e.\ in the encoded model where the gauge fields have been analytically eliminated) is identical to the half-chain entropy in the original model (which includes both matter- and gauge fields). This is a consequence of  the fact that the gauge degrees of freedom are fixed for a certain spin configuration and background field in combination with the requirement that only states within a fixed charge sector are considered.

\begin{figure}[t]
\centering
\includegraphics[width=0.85\textwidth, angle=0]{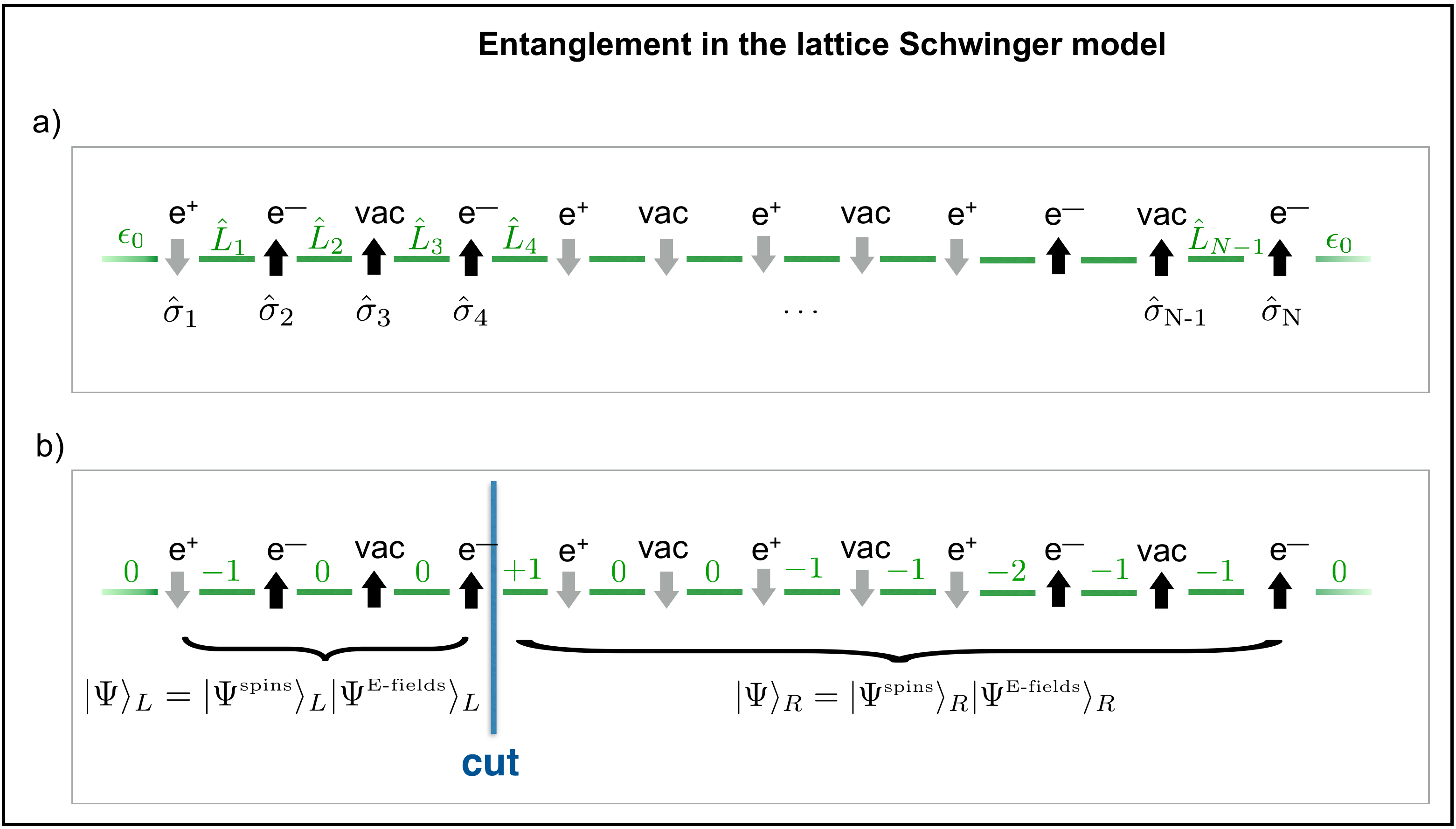}
\caption{\label{Fig_Entanglement} Illustration of a quantum state that respects Gauss' law. As explained in the text, we introduce a bipartiton that is defined by a cut dividing the system into a left part and a right part. For analyzing the entanglement between these parts, we consider superpositions of states of the shown type, where the left (right) part of the system is described by $|\Psi\rangle_{\text{\tiny{L/R}}}=|\Psi^{\text{\tiny{spins}}}\rangle_{\text{\tiny{L/R}}}|\Psi^{\text{\tiny{E-fields}}}\rangle_{\text{\tiny{L/R}}}$. }
\end{figure}

As explained in detail in section~\ref{SubSec_Encoding}, the encoding of the Schwinger model in a pure-spin Hamiltonian involves a gauge transformation that eliminates the gauge field operators which represent the vector potentials, as well as a Jordan--Wigner transformation that maps fermionic degrees of freedom to spin operators. Neither of these transformations has an effect on the correlations between the left and right half of the system (this can be understood by noting that either transformation acts on the left subsystem only with operators that are entirely contained within this part of the system). In the following, we discuss therefore the last step of the encoding, which entails the elimination of the gauge field operators that represent the electric fields in the continuum limit. We consider the situation illustrated in Fig.~\ref{Fig_Entanglement}a, which involves spin operators $\hat{\sigma}_n$ at lattice sites $n$ and electric field operators $\hat{L}_n$  that are defined on the links between two lattice sites and take integer eigenvalues $L_n=0,\pm1,\pm2,...$. We assume that the background electric field takes the value $\epsilon_0=0$ (as in section~\ref{SubSec_Encoding}) and consider a bipartition that is defined by a cut as shown in Fig.~\ref{Fig_Entanglement}b. The Gauss law $\hat{L}_n-\hat{L}_{n-1}=\frac{1}{2}\left(\hat{\sigma}_n^z+(-1)^n\right)$ can be used to determine the values the electric field stepwise from left to right (compare Fig.~\ref{Fig_1}, panels b and c), such that the state of the full quantum state of the system can be written in the form
\begin{eqnarray*}
|\Psi\rangle=\sum_{ij} \lambda_{ij}\ \! |\Phi_{i}^{\text{\tiny{spins}}}\rangle_L|\Phi^{\text{\tiny{E-fields}}}_i\rangle_L\otimes   |\Phi_{j}^{\text{\tiny{spins}}}\rangle_R|\Phi^{\text{\tiny{E-fields}}}_{ij}\rangle_R,
\end{eqnarray*}
with coefficients $\lambda_{ij}$. In this notation, the subscript L (R) refers to the left (right) part of the chain with respect to the chosen cut. For each side, the quantum state can be written as a tensor product of a spin state and a state that refers to the gauge fields. Note that the state referring to the gauge fields of the right part of the system depends on the spin- (and corresponding gauge field-) state of the left side, as indicated by the subscripts of the last ket. However, due to charge conservation, the Gauss law can also be used to determine the electric fields starting from the right end of the chain, where the background field also takes the value $\epsilon_0=0$. Therefore, the full quantum state of the system can be expressed in the form
%
%
%
%
%
\begin{eqnarray*}
|\Psi\rangle=\sum_{ij} \lambda_{ij}\  \! |\Phi_{i}^{\text{\tiny{spins}}}\rangle_L|\Phi^{\text{\tiny{E-fields}}}_i\rangle_L\otimes   |\Phi_{j}^{\text{\tiny{spins}}}\rangle_R|\Phi^{\text{\tiny{E-fields}}}_{j}\rangle_R.
\end{eqnarray*}
This is a consequence of the fact that the electric fields of the right side can be inferred from the background field and spin configuration on the right side alone. The gauge field at the cut can be inferred by measuring the spins on either the right or the left side separately. In other words, the gauge field at the link bears no additional information once the spin state on either of the two sides is known. This applies also to the gauge field at the cut. Hence, the entanglement between the two parts of the system can be calculated using the reduced spin state
\begin{eqnarray*}
|\Psi'\rangle=\sum_i \lambda_{ij} \ \!|\Phi_{i}^{\text{\tiny{spins}}}\rangle_L\otimes   |\Phi_{j}^{\text{\tiny{spins}}}\rangle_R.
\end{eqnarray*}
It does not matter where the cut is placed with respect to the position to the nearest gauge field. The entanglement can be evaluated by including the gauge field nearest to the cut to either side.
Similarly, methods that are based on doubling the link and the corresponding gauge field and assigning one copy to either part of the chain~\cite{Casini2014} lead to the same result. This argument can be straightforwardly generalized  to mixed states.

We note that the entanglement with respect to the original model is not given by the entanglement in the reduced spin system if the requirement of charge conservation is not fulfilled. In this case, there exist for example separable reduced spin states which correspond to entangled states of the full system (involving both spin- and gauge fields). However, this is not a  concern as long as charge conservation can be guaranteed to be respected by the considered time evolution. In the Schwinger model, changes in the fermion number are only possible through charge conserving particle-antiparticle creation or annihilation events. This is also the case for the discretized dynamics realized in our simulation scheme (Trotter errors do not violate charge conservation). Implementation errors such as spin flips (which correspond to the creation or annihilation of a single charge) can lead to states that do not respect charge conservation. In this case, the total magnetisation of the spin system is nonzero, and the corresponding states can be detected and filtered out by postselection (see Sec.~\ref{SubSec_Decoupling}).

\medskip

\bibliographystyle{iopart-num}
\bibliography{bib}

\end{document}